\documentclass[submission, Phys]{SciPost}

\hypersetup{
    colorlinks,
    linkcolor={red!50!black},
    citecolor={blue!50!black},
    urlcolor={blue!80!black}
}

\usepackage[bitstream-charter]{mathdesign}
\DeclareSymbolFont{usualmathcal}{OMS}{cmsy}{m}{n}
\DeclareSymbolFontAlphabet{\mathcal}{usualmathcal}

\usepackage{graphicx}
\usepackage{svg}

\usepackage{upgreek} 

\usepackage{amsmath}
\usepackage{color}

\usepackage{amsmath}
\usepackage{soul}

\newcommand{\dd}{\mathrm{d}}   \newcommand{\dr}{\partial}   \newcommand{\ee}{\mathrm{e}}  \newcommand{\ii}{\mathrm{i}}  \newcommand{\pii}{\uppi}  \newcommand{\Ga}{\varGamma} \newcommand{\La}{\varLambda} \DeclareMathOperator{\tr}{\mathrm{Tr}}  \newcommand{\fig}{Fig.{}}  \newcommand{\kf}{k_{\mathrm{F}}} \newcommand{\sgn}{\mathrm{sgn}\,} 

\newcommand{\Gav}{\boldsymbol{\varGamma}}     \newcommand{\Rv}{\boldsymbol{R}}  \newcommand{\Gv}{\boldsymbol{G}}   \newcommand{\Sv}{\boldsymbol{S}}

\newcommand{\FD}{n_{\textsc{f}}}

\newcommand{\vf}{v_{\mathrm{F}}}  \newcommand{\derr}{\delta^{\textsc{rl}}}

\newcommand{\ehm}{EHM{}}

\begin{document}

\begin{center}{\Large \textbf{
Functional renormalization group for fermions on a one-dimensional lattice at arbitrary filling\\
}}\end{center}

\begin{center}
L. Désoppi\textsuperscript{1, 2$\star$},
N. Dupuis\textsuperscript{2} and
C. Bourbonnais\textsuperscript{1}
\end{center}

\begin{center}
{\bf 1} Regroupement Québécois sur les Matériaux de Pointe and Institut Quantique, \\
Département de physique, Université de Sherbrooke, \\
Sherbrooke, Québec, Canada, J1K-2R1
\\
{\bf 2} Sorbonne Université, CNRS, Laboratoire de Physique Théorique de la Matière Condensée,
LPTMC, F-75005 Paris, France
\\
${}^\star$ {\small \sf Lucas.Desoppi@USherbrooke.ca}
\end{center}

\begin{center}
\today
\end{center}


\section*{Abstract}
{\bf
A formalism based on the fermionic functional-renormalization-group approach to interacting electron models defined on a lattice is presented. One-loop flow equations for the coupling constants and susceptibilities in the particle-particle and particle-hole channels are derived in weak-coupling conditions. It is shown that lattice effects manifest themselves through the curvature of the spectrum  and  the dependence of the coupling constants on momenta.  This method is then applied to the one-dimensional extended Hubbard model; we thoroughly discuss the evolution of the phase diagram, and in particular the fate of the bond-centered charge-density-wave phase, as the system is doped away from half-filling. Our findings are compared to the predictions of the field-theory continuum limit and available numerical results.}

\vspace{10pt}
\noindent\rule{\textwidth}{1pt}
\tableofcontents\thispagestyle{fancy}
\noindent\rule{\textwidth}{1pt}
\vspace{10pt}

\section{Introduction}

 The theory of interacting fermions in one spatial dimension  gives the best understood examples of models  whose asymptotic low-energy behavior distinctively deviates from that of a Fermi liquid, as commonly found in Fermi systems  in higher dimension. Absence of quasi-particle excitations and power-law decay of correlation functions   are  governed by  non-universal exponents characterized by  very  few   hydrodynamic and interaction-dependent parameters which separate into  spin and charge bosonic entities  for spin-$\frac{1}{2}$ fermions \cite{Emery79,Solyom79,Schulz95,Voit95,Bourbon91,Giamarchi04}. Such  distinctive features form the basis  of the Luttinger liquid (LL) fixed-point phenomenology \cite{Haldane81}. This is  asymptotically accurate in the low-energy (continuum) limit, namely  when the fermion spectrum can be considered as strictly linear around the Fermi points and  when interactions  projected on those points are considered  as momentum independent.  These  are well known to be at the core of  the  field theory or continuum $g$-ology models of the interacting 1D  Fermi gas.   The fixed-point behavior of a linear LL proves to be generic for gapless branches of excitations of most models of interacting fermions in  one dimension.

As one moves  away in energy from the Fermi points  the  spectrum develops in practice some  curvature. Deviations with respect to linearity alongside  momentum dependence of interactions, although irrelevant in the renormalization group (RG) sense \cite{Haldane81},  were shown to modify the finite energy  spectral properties  predicted by the   linear LL theory.   Formulated in terms of an effective  x-ray edge problem \cite{Pustilnik06}, the coupling of particles  to a  continuum of higher energy states is found to  alter the power-law profiles of spectral lines  near their absorption edges\cite{Markhof16}. These non-linear LL effects could be rigorously checked  in the case of  integrable spinless-fermion models defined  on a lattice \cite{Pereira09,Imambekov12,Cheianov08,Meden19}. 

Noticeable limitations of the linear $g$-ology mappings of non-integrable lattice models  could also be found in the calculation  of  singular correlations that enter in the determination of their phase diagrams. This has been best exemplified   in the case of the one-dimensional extended Hubbard model (EHM) for spin-$\frac{1}{2}$  fermions, which will serve here as the reference lattice model for the RG method developed in the present work.

At half-filling numerical calculations soon  identified a shift of the continuous transition line connecting charge- and spin-density-wave states \cite{Fourcade84,Hirsch84}, a line  that the continuum $g$-ology theory predicts to be gapless   along the separatrix $U=2V$,  for the local ($U$) and nearest-neighbors ($V$) interaction parameters  of the EHM. The origin of this alteration  has resisted  at least in weak coupling to all attempts  of explanations  formulated in the framework of the linear $g$-ology theory \cite{Voit92,Cannon90}.  Using exact diagonalizations, Nakamura  showed later on  that the shift underlies the incursion of a distinct  phase, known as a bond-centered charge-density-wave (BOW) phase. The BOW phase  is  entirely gapped  in both spin and charge sectors and   extends across some finite region  on both sides of the $U=2V$ line of the  phase diagram in weak coupling   \cite{Nakamura99,Nakamura00}. This was  subsequently confirmed numerically both by quantum Monte Carlo \cite{Sengupta02,Sandvik04}, and density-matrix RG  methods \cite{Zhang04,Jeckelmann02,Ejima07}. 

On analytical grounds, Tsuchiizu and Furusaki showed from perturbation theory that by taking into account   the momentum-dependent fermion-fermion  scattering processes at high energy, that is,   beyond the linear region,  one can define, at some  arbitrarily chosen lower energy, an effective weak-coupling linear $g$-ology  model, but with modified and enlarged set of input parameters\cite{Affleck}. The modification is such that it allows  the emergence of a BOW phase in the $U=2V$ weak-coupling sector of the phase diagram \cite{Tsuchiizu02,Tsuchiizu04}. Using a  functional fermionic RG approach at the one-loop level, Tam {\it et al.}\cite{Tam06} pointed out that by integrating out numerically all the scattering processes  for  a discrete set of fermion momenta along the  tight-binding spectrum  in  the  Brillouin zone, the existence of a BOW phase  can be found in the $U=2V$ weak-coupling region  of  the  EHM phase diagram at half-filling. Ménard  {\it et al.}\,\cite{Menard11} thereafter formulated an RG transformation    for half-filling  tight-binding fermions in the  Wilsonian   scheme \cite{Dumoulin96},  in which irrelevant interaction terms can be classified from the  momentum dependence of  non-local scattering amplitudes away from the Fermi surface. Their impact  on the low-energy RG flow has born out the presence of the BOW ordered phase where it is expected in the EHM phase diagram at weak coupling, alongside shifts of some other transitions lines where accidental symmetries are known to occur in the continuum $g$-ology limit.   

 These RG results focused on the EHM model at half-filling. It is the main motivation of  the present  work to extend  away from half-filling the weak-coupling  RG  determination of quantum phases of 1D fermion lattice models in the presence of  non local interaction. To achieve this program, we shall opt for the functional RG approach in the so-called one-particle-irreductible scheme \cite{Metzner12,Dupuis21}, which proves  easier to implement analytically when dealing with the relatively unexplored situation of momentum dependent interactions and asymmetrically  filled tight-binding spectrum. The one-loop RG flow equations for the momentum-dependent four-point vertices are expanded up to second order in the energy difference from the Fermi level for the asymmetric  spectrum. The difference acts as the scaling variable which allows the  power counting classification of marginal and irrelevant interaction terms, together with their interplay.  The method developed below can in principle apply to any form of non-linear spectrum  and momentum-dependent interactions in models  with  fermion density away from  half-filling.  From the calculations of the most singular susceptibilities the phase diagram of the EHM model can be mapped out. At half-filling the results confirm previous RG calculations for the existence of a gapped BOW phase near the $U=2V$ line and bear out  the shift of other transition lines between different ground states, in agreement with numerical results \cite{Nakamura00}. In both situations the role of the spectrum and irrelevant interactions terms in the qualitative change of initial conditions for an  effective linear continuum theory in the low-energy limit can be established.  The method is carried out   away from half-filling and the region of dominant BOW gapped  state is found to gradually shrink in size to ultimately be suppressed as a function of doping. The whole phase diagram  then evolves towards  an incommensurate situation but where noticeable modifications of the stability regions of quantum states, as predicted by the  $g$-ology continuum model,  are found. The  integration of  high-energy electronic states in the  particle-hole-asymmetric non-linear part of the spectrum reveals the existence of screening effects coming from particle-particle pairing fluctuations   which act at lower energy as an important  factor in promoting   singlet  superconductivity or inversely  either antiferromagnetism or triplet superconductivity  against the charge-density-wave state.      
 
 The paper is organized as follows. In Sec.\,II the fRG method is introduced and the flow equations of couplings and various susceptibilities are derived at the one-loop level. In this framework known results  of the EHM phase diagram in the limit of the continuum $g$-ology model at and away from half-filling are recovered. In Sec.\,III, we  broaden the formulation of fRG  to include the tight-binding  spectrum and the momentum-dependent interactions of the EHM, as actually defined on a lattice. The one-loop flow equations for marginal and  up to second order for the   set of irrelevant scattering amplitudes are derived. The phase diagrams  at and away from half-filling   are obtained and  their  comparison with the $g$-ology limit  analyzed and critically discussed. We conclude this work in Sec.~\ref{conclusion}.

\section{FRG for the extended Hubbard Model at arbitrary filling}

\subsection{One-dimensional Extended Hubbard Model}

The 1D extended Fermi-Hubbard model  is defined  by the Hamiltonian (in this paper, units are taken such that $ k_{\mathrm{B}} = \hbar = 1 $ and the lattice constant $a=1$)
\begin{equation}
\mathcal{H}  =  - t \sum_{i, \, \sigma} \big(c^{\dagger}_{i, \, \sigma} c^{\phantom{\dagger}}_{i+1, \, \sigma} + \mathrm{H.c.} \big)  
 +  U \sum_{i} n_{i, \, \uparrow} n_{i, \, \downarrow}  +  V \sum_{i}  n_{i} n_{i+1},
\label{ham}
\end{equation} 
describing electrons moving on a lattice with a hopping amplitude $t>0$ and experiencing on-site and nearest-neighbor interactions with strengths $U$ and $V$, respectively. In Eq.~(\ref{ham}), $i$ denotes the site index,  $\sigma = \uparrow, \downarrow $ is the spin index, $ n_{i, \, \sigma} = c^\dagger_{i, \, \sigma}  c_{i, \, \sigma}$ and $ n_i = n_{i, \uparrow} + n_{i, \downarrow} $ is the number of electrons at site $i$. 

\begin{figure}
    \centering
    \includegraphics[scale = 0.5]{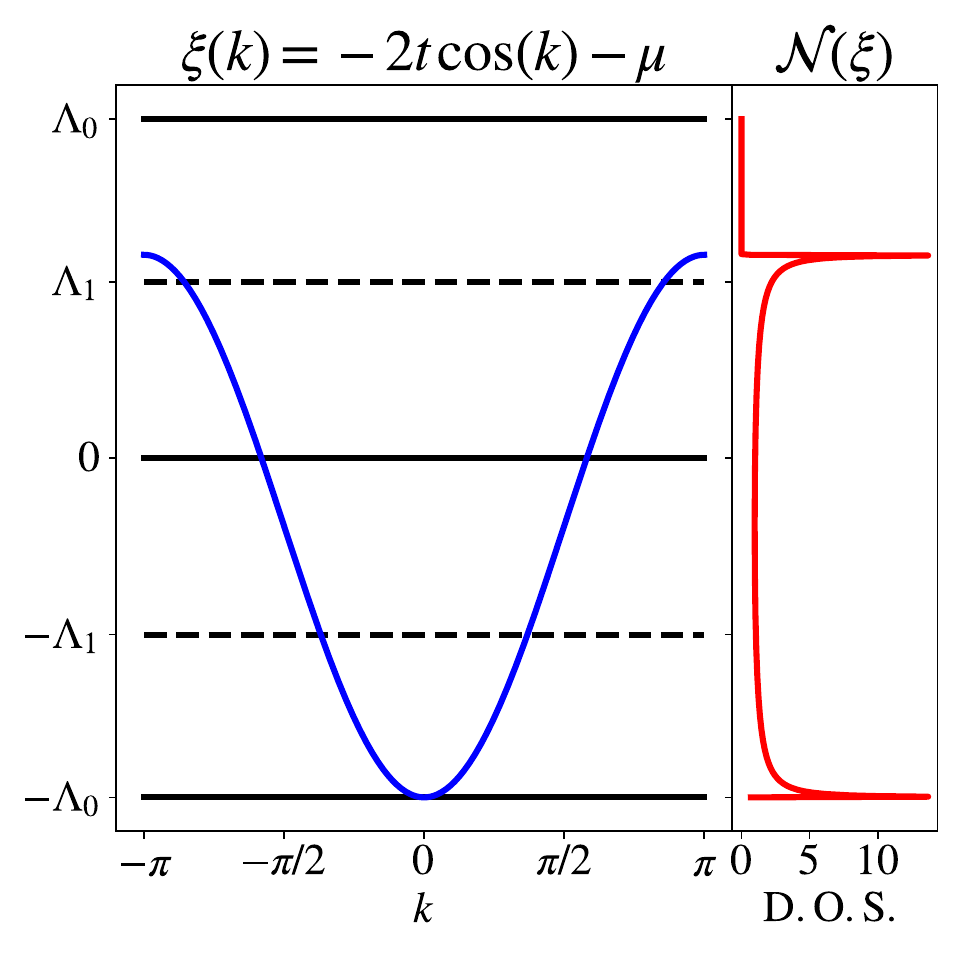}
    \caption{Tight-binding spectrum of the EHM model. Here $\La_0$ is half of the initial bandwidth ($ \La_0 = 2 t + | \mu |  $) and $\La_1$ is the energy cutoff at some intermediate step of the RG flow. On the right panel, $\cal{N}(\xi)$ the density of states as a function of energy showing the van-Hove singularities at the band edges.}  .
    \label{TightBinding}
\end{figure}

The one-particle states have energies $ \varepsilon(k) = -2 t \cos(k) $ with wave vector $k$ of the tight-binding form, such that  with respect to the Fermi level, these are comprised in the interval  $  -2 t - \mu \leqslant \xi = \varepsilon - \mu \leqslant 2 t -\mu   $, where $\mu $ is the  chemical potential. The tight-binding spectrum $\varepsilon(k)$ is shown in \fig \ref{TightBinding}. The corresponding density of states is written as follows:

\begin{equation}
\mathcal{N}(\xi) = \dfrac{  \Theta(2 t - |\mu + \xi|)} {2\pii \sqrt{t^2 - (\xi+\mu)^2/4 } },
\end{equation}
where $ \Theta(x) $ is the Heavisde step function. It will  indeed be useful to write the density of states for an arbitrary value of $ \xi $, because in the RG flow, the momentum shell corresponding to the integration of the degrees of freedom will be taken at equal distance from the Fermi level for the empty and the occupied states (see Fig.~\ref{TightBinding}). By definition, the Fermi level is related to the Fermi wave vector $\kf$, defined such that $ \varepsilon(\kf) = \mu $. One can also define the Fermi velocity
\begin{equation}
    \vf = \left.\dfrac{\dr \varepsilon}{\dr k}\right|_{\kf} = 2 t \sin(\kf).
\end{equation}
Let $n$ be the fermion  filling number. Obviously, we have $ 0 \leqslant n \leqslant 2 $. This number is directly given by an integration of the density of states up to the Fermi level:
\begin{equation}
    n = 2\int\limits_{-\infty}^0 \dd \xi \, \mathcal{N}(\xi) = 2 \int\limits_{-\kf}^{+\kf} \dfrac{\dd k}{2\pii} = \dfrac{2 \kf }{\pii},
\end{equation}
which leads to the simple relations: 
\begin{equation}
\label{mu}
    \kf = \dfrac{\pii }{2} n, \qquad  \mu = - 2  \cos( \pii n /2),
\end{equation}
where from now on $\mu$ is expressed in units of $t$. 

 In reciprocal space, the Hamiltonian of the EHM is written as (where $L$ denotes the number of lattice sites)
\begin{align}
    H & = \sum_{ k, \, \sigma  } \big( \varepsilon(k) - {U/2} \big) c^{{\dagger}}_{k, \, \sigma} c^{\phantom{\dagger}}_{k, \, \sigma} 
     +  \dfrac{\pii \vf}{2 L}  \sum_{\{ k, \, \sigma \}}  g_{k_1, \, k_2, \, k'_{1}}     c^\dagger_{k'_{1}, \, \sigma_1}  c^\dagger_{k'_{2}, \, \sigma_2}  c^{\phantom{\dagger}}_{k_2, \, \sigma_2}  c^{\phantom{\dagger}}_{k_1, \, \sigma_1}  
    \derr_{k_1+k_2-k'_{1}-k'_{2}},
\end{align}
where $ \derr $ denotes the momentum conservation condition on the lattice (RL stands for Reciprocal Lattice): 
\begin{equation}
\derr_{k} = \sum_{n = -\infty}^{+\infty} \delta_{k, 2\pii n},
\end{equation}
and the dimensionless coupling constants are given by:
\begin{equation}
g_{k_1, \, k_2, \, k'_{1}} = \dfrac{U}{\pii \vf } + \dfrac{2 V}{\pii \vf} \cos(k_1 -k'_{1} ).
\label{gInit}
\end{equation}

\subsection{One-loop flow equations}

The EHM is studied with the functional RG.   We  first recast the partition function of the model into a field-theory setting at finite temperature $ T = 1/\beta $, by means of a functional integral over a Grassmannian field    $\varphi$:
\begin{equation}
\mathcal{Z} = \tr \ee^{-\beta ( \mathcal{H} - \mu \mathcal{N} )} = \int \mathcal{D}[\varphi] \ee^{- \mathcal{S}[\varphi] } ,
\end{equation}
where the action $ \mathcal{S}[\varphi] $ takes the  form
\begin{equation}
    \mathcal{S}[\varphi] = - \sum_{a',a,\sigma}\ \big[ G^0 \big]^{-1}_{a'a}  \bar\varphi_{a', \sigma} \varphi_{a, \sigma} +  \dfrac{ T}{2 L}\sum_{\{a',a,\sigma\}} V_{a'_1 a'_2 a_2 a_1}   \bar\varphi_{a'_1, \sigma_1} \bar\varphi_{a'_2, \sigma_2} \varphi_{a_2,\sigma_2} \varphi_{a_1, \sigma_1}
\end{equation}
and the index $a \rightarrow (\omega_n, k)$ carries all the relevant information about momentum $k$ and fermionic Matsubara frequency $ \omega_n = (2n+1)\pii T $. 

The first term in the action is related to the free propagator $ G^0 $ which is diagonal in reciprocal space
\begin{equation}
  \big[ G^0 \big]^{-1}_{a'a} = \big(\ii \omega_n - \xi(k) \big) \delta_{a'a}. 
\end{equation}
The second term describes two-body interactions, and takes the following form:
\begin{equation}
  V_{a'_1 a'_2 a_2 a_1  } = \pii \vf   g_{k_1, \, k_2, \, k'_1}  \derr_{k'_1 + k'_2 - k_2 - k_1}\delta_{\omega_{n'_1}+\omega_{n'_2}-\omega_{n_2}-\omega_{n_1}, 0}.  
\end{equation}

\begin{figure}
    \centering
    \begin{align*}
\La\dr_\La g  & = \vcenter{\hbox{\includegraphics[scale=0.32]{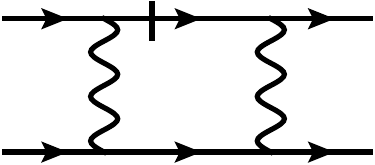}}} + \raisebox{-11.75pt}{\includegraphics[scale=0.32]{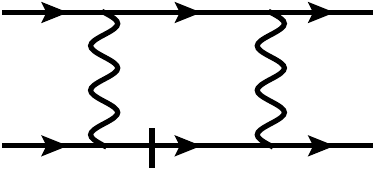}}  +  \vcenter{\hbox{\includegraphics[scale=0.32]{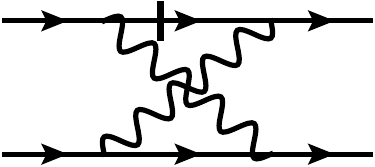}}} + \raisebox{-11.75pt}{\includegraphics[scale=0.32]{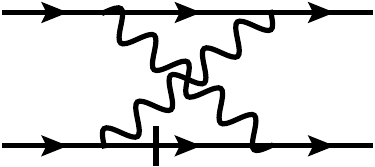}}  \\
 & \ \ +  \vcenter{\hbox{\includegraphics[scale=0.32]{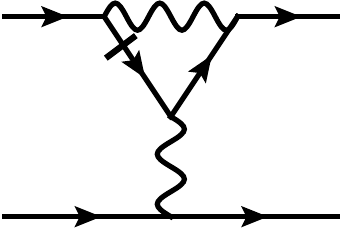}}} + \vcenter{\hbox{\includegraphics[scale=0.32]{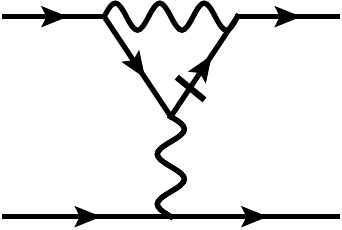}}} + \vcenter{\hbox{\includegraphics[scale=0.32]{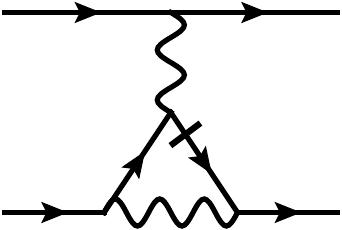}}} + \vcenter{\hbox{\includegraphics[scale=0.32]{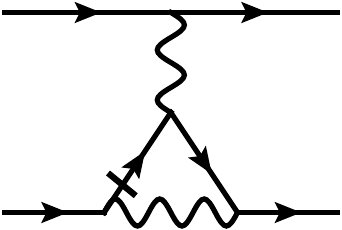}}} \\
  & \ \ +  \vcenter{\hbox{\includegraphics[scale=0.32]{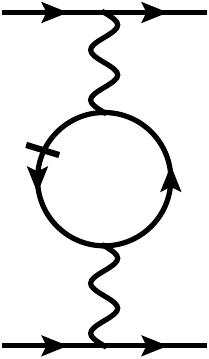}}} + \vcenter{\hbox{\includegraphics[scale=0.32]{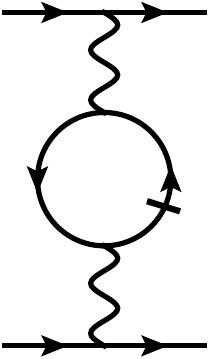}}}  
\end{align*}
    \caption{One-loop flow equations of the coupling constants in diagrammatic form. Here a slashed line refers to the single-scale propagator. 
    }
    \label{dgDiag}
\end{figure}

\begin{figure}
    \centering
    \begin{align*}
\La\dr_\La Z^{\mathrm{ch/sp}} & = \raisebox{-19pt}{\includegraphics[scale=0.32]{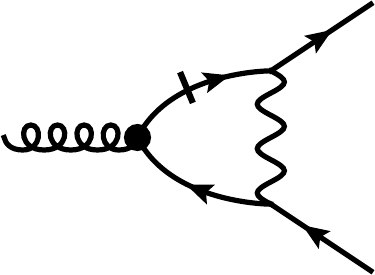}} + \raisebox{-9pt}{\includegraphics[scale=0.32]{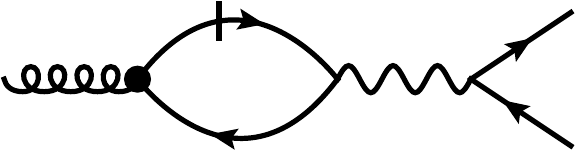}}  \nonumber \\
& \ \  + \raisebox{-19pt}{\includegraphics[scale=0.32]{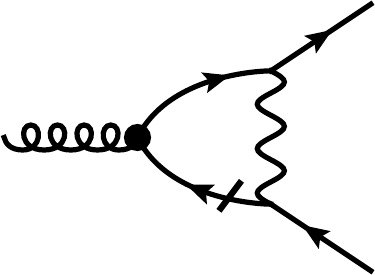}}+ \raisebox{-9pt}{\includegraphics[scale=0.32]{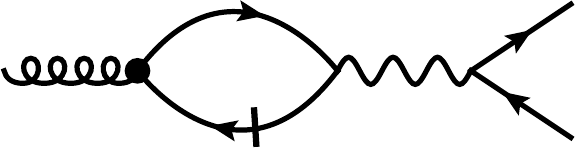}}  \\
\La\dr_\La Z^{\mathrm{s/t}}  & = \raisebox{-19pt}{\includegraphics[scale=0.32]{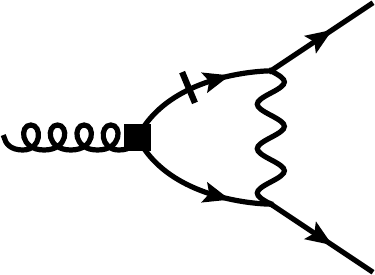}} + \raisebox{-19pt}{\includegraphics[scale=0.32]{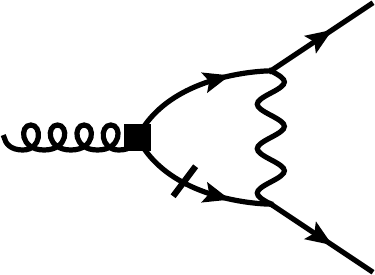}} 
\end{align*}
    \caption{One-loop corrections the flow equations of three-leg  vertices for charge/spin-density-wave and singlet/triplet-pairing susceptibilities.} 
    \label{dZDiag}
\end{figure}

\begin{figure}
    \centering
    \begin{align*}
\La\dr_\La \chi^{\mathrm{ch/sp}} & = \raisebox{-9pt}{\includegraphics[scale=0.32]{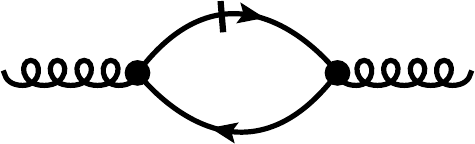}} + \raisebox{-9pt}{\includegraphics[scale=0.32]{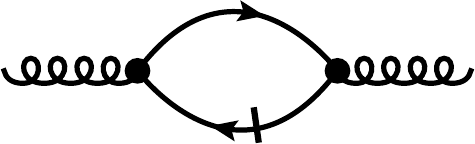}}  \\
\La\dr_\La \chi^{\mathrm{s/t}} & = \raisebox{-9pt}{\includegraphics[scale=0.32]{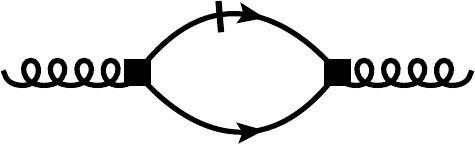}}  + \raisebox{-9pt}{\includegraphics[scale=0.32]{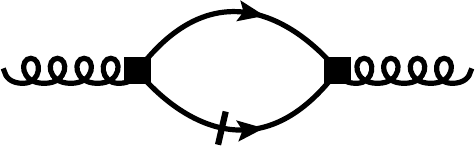}}
\end{align*}
    \caption{ Flow equations for the charge/spin-density-wave and singlet/triplet-superconducting  susceptibilities. }
    \label{dChiDiag}
\end{figure}

A quadratic term is added to the action, 
\begin{equation}
    \mathcal{S}[\varphi] \rightarrow \mathcal{S}[\varphi] + \sum_{a',a,\sigma} \bar\varphi_{a', \sigma} R_{\La, a' a} \varphi_{a, \sigma},
\end{equation}
which regularizes the functional integral by suppressing the low-energy fluctuations. An anticommuting source  field $\eta,\bar{\eta}$ coupled to the fermion field is also included in the action  which takes the form $ \sum_{a,\sigma}\big( \bar\eta_{a,\sigma}  \varphi_{a,\sigma}+ \bar\varphi_{a,\sigma}\eta_{a,\sigma}  \big) $. This  gives the regularized generating functional of  correlation functions $ \mathcal{Z}_{\La}[\eta] $. The regularized effective action  $ \Ga_{\La}[\phi] $ is then defined as the modified Legendre transform of the generating functional of  connected correlation functions {$ \mathcal{W}_{\La}[\eta]$   $ = \log \mathcal{Z}_{\La}[\eta] $}:
\begin{equation}
    \Ga_{\La}[\phi] + \mathcal{W}_{\La}[\eta] =  \sum_{a,\sigma}  \big(\bar\eta_{a,\sigma}   \phi_{a,\sigma} + \bar \phi_{a,\sigma}\eta_{a,\sigma}   \big)-  \sum_{a',a,\sigma}  \bar\phi_{a', \sigma} R_{\La, a' a} \phi_{a, \sigma} ,
\end{equation}
where $\phi_{a,\sigma}=\langle \varphi_{a,\sigma}\rangle$ and $\bar\phi_{a,\sigma}=\langle \bar\varphi_{a,\sigma}\rangle$ with the expectation values computed in the presence of the source fields $\eta,\bar\eta$. The regularized effective action $\Ga_\La[\phi]$ satisfies
the Wetterich equation\cite{Wetterich93,Ellwanger94,Morris94} 
\begin{equation}
\dr_\La\Ga_\La[\phi] =  \frac{1}{2} \tr \, \left\{ \dr_\La \Rv_{\La } \left( \Gav_\La^{(2)}[\phi] + \Rv_{\La } \right)^{-1} \right\}, 
\end{equation}
where $ \Gav_\La^{(2)}[\phi]  $ is the second functional derivative of the effective action with respect to the field.
Additional source fields $J$ can be added to the effective action in order to generate flow equations for the response functions. The idea is then to decompose $ \Ga_\La[\phi, J] $ as a sum of monomials $ \Ga^{[n, p]}_\La [\phi, J] \sim \phi^n J^p $, and make identifications on both sides of the flow equation. We proceed at the  one-loop level for which the 1-PI fRG hierarchy is truncated,  this procedure leads to flow equations in weak coupling for the coupling constants $g$,   three-leg vertices $Z$ and  susceptibilities $\chi$. These equations have the familiar schematic form
\begin{equation}
    \La\dr_\La g \sim \int \mathcal{L} g g, \quad \La\dr_\La Z \sim \int \mathcal{L} Z g, \quad \La\dr_\La \chi \sim \int \mathcal{L} Z Z,
\end{equation}
where $\La$ selects the degrees of freedom that are integrated at the step $ \La $.  The corresponding one-loop diagrammatic contributions to the flow equations  are  shown in Figs.~\ref{dgDiag}, \ref{dZDiag} and \ref{dChiDiag}, respectively, and where a simple line corresponds to the propagator $ \Gv_\La = \left( \Gav_\La^{(2)}[\phi = 0] + \Rv_{\La } \right)^{-1} $ and a slashed line to the single-scale propagator $ \Sv_\La = - \Gv_\La \dr_\La\Rv_\La \Gv_\La $.

\subsection{Recovery of the $g$-ology continuum model}

Before we take into account the non-linearity of the spectrum and the irrelevant coupling constants, it is useful for later comparisons to recover the well known  $g$-ology  electron gas model in the continuum limit, also known as the 1D electron gas model, for which lattice effects are mostly discarded. Thus,  we linearize the tight-binding spectrum $ \xi(k) = \varepsilon(k) - \mu $ in the vicinity of the two Fermi points $\pm \kf $. We can write $ k = \eta \kf + ( k - \eta \kf ) $, where $ \eta = \pm1 $ is the branch index, which gives 
\begin{equation}
\xi(k) =  \left. \dfrac{\dr \varepsilon}{\dr k} \right|_{\eta \kf} (k  -  \eta \kf  ) + \dots \, =\, \vf(|k| -  \kf) + \dots \text{.}
\end{equation}
In  the particular case of a half-filled band,   $ \kf = \pii/2 $. The reciprocal lattice wavevector $G=2\pii$ being equal to $4\kf$, we can identify $\pm 3\kf$ with $\mp \kf$. Away from half-filling, $G\neq 4\kf$ and this identification does no longer hold.  The $g$-ology model  is obtained when the momenta appearing in the coupling constants are evaluated on the Fermi points
\begin{alignat}{3}
\label{gini}
 g_{1} & \equiv  g_{+\eta\kf, \,  -\eta\kf, \,  -\eta\kf} && = \big(U - 2 V( 1 - \mu^2/2  )\big)/\pii \vf, \qquad && \cr
 g_{2} & \equiv  g_{+\eta\kf, \,  -\eta\kf, \,  +\eta\kf} &&= (U + 2 V)/\pii \vf , &&  \cr
 g_{3} & \equiv  g_{+\eta\kf, \,  +\eta\kf, \,  -\eta\kf} &&=  \big(U - 2 V( 1 - \mu^2/2  )\big)/\pii \vf, &&  \cr
 g_{4} & \equiv  g_{+\eta\kf, \,  +\eta\kf, \,  +\eta\kf} && =  (U + 2 V)/\pii \vf, && 
\end{alignat}
where the initialization condition \eqref{gInit} has been used. It appears that the constants $ g_2 $ and $g_4$ correspond to forward scattering, $g_1$ to backward scattering, while $g_3$ describes umklapp processes \cite{Dzyaloshinskii72}. In order to find the expression of the two-particle vertex $ \Ga^{[4,0]}[\phi] $, we write its restriction close to the two Fermi points, which is indicated by the bracket $ [ \cdot ]_{\mathrm{F}} $: 
\begin{align}
\Big[ \Ga^{[4,0]}[\phi]  \Big]_{\mathrm{F}} 
={}& \dfrac{\pii \vf T}{L} \sum_{\eta, \, \sigma_i} \sum_{\omega_{n_i}} \sum_{k_i\geqslant 0} \bigg\{ \frac{1}{2}( g_1 \delta_{\sigma_1, \, \sigma_3} \delta_{\sigma_2, \, \sigma_4}   - g_2 \delta_{\sigma_1, \, \sigma_4} \delta_{\sigma_2, \, \sigma_3} )  \nonumber \\
& \hspace{3.45cm} \times \bar\phi_{-\eta k'_1, \, \sigma_3} \bar\phi_{+\eta k'_2, \, \sigma_4}\phi_{-\eta k_2, \, \sigma_2}\phi_{+\eta k_1, \, \sigma_1} \nonumber \\ & \hspace{2.75cm} 
 + \dfrac{ g_3}{2 }   \bar\phi_{+\eta k'_1, \, \sigma_1} \bar\phi_{+\eta k'_2, \, \sigma_2}\phi_{-\eta k_2, \, \sigma_2}\phi_{-\eta k_1, \, \sigma_1} \nonumber \\ 
& \hspace{2.75cm}  +\dfrac{ g_4}{2 }    \bar\phi_{+\eta k'_1, \, \sigma_1} \bar\phi_{+\eta k'_2, \, \sigma_2}\phi_{+\eta k_2, \, \sigma_2}\phi_{+\eta k_1, \, \sigma_1} \bigg\}  ,
\end{align}
where momentum and Matsubara-frequency conservation is understood in the right-hand side.

When this simplified vertex is inserted in Eqs.~\eqref{FlowG} (see Appendix~\ref{Appendix}), we end up with the well known $g$-ology flow equations
\begin{equation}
\begin{split}
\label{gologyg}
\La\dr_\La g_1 & = \mathcal{L}_{\mathrm{P}} g_1^2 -  (\mathcal{L}_{\mathrm{C}}+\mathcal{L}_{\mathrm{P}}) g_1 g_2 -  \mathcal{L}_{\mathrm{L}} g_1 g_4 ,  \\
\La\dr_\La  g_2  & = - \mathcal{L}_{\mathrm{C}} g^2_1/2 -    (\mathcal{L}_{\mathrm{C}}+\mathcal{L}_{\mathrm{P}}) g^2_2 /2  
-   \mathcal{L}_{\mathrm{P}'}    g^2_3 /2 -  \mathcal{L}_{\mathrm{L}} g_4 (g_1 - 2 g_2), \\
\La\dr_\La g_3  & =    (\mathcal{L}_{\mathrm{P}} + \mathcal{L}_{\mathrm{P}'})   g_3 ( g_1 - 2g_2 )/2  
- \mathcal{L}_{\mathrm{C}'}   g_3(g_2 + g_4)/2, \cr
\La\dr_\La  g_4  & = - \mathcal{L}_{\mathrm{L}} ( g^2_1 - 2 g^2_2 + 2 g_1 g_2 + g^2_4  )/2  
- \mathcal{L}_{\mathrm{C}'} (   g^2_3 + g^2_4 )/2.
\end{split}
\end{equation}
Here the $\mathcal{L}_{\mathrm{X}}$'s are derivatives with respect to $\varLambda$ of the  bubbles associated to particle-particle (p-p) and particle-hole (p-h) scattering channels    in which  $\mathrm{C}$ and $\mathrm{C}'$ refer to inter- and intra-branch Cooper pairings, and   $\mathrm{P}$ and $\mathrm{L}$ refer to Peierls and Landau channels.     For this particular calculation a sharp cutoff is chosen (see appendix \ref{Regul} concerning the regulator). This allows to compute the integrals in closed form and to recover the known results of the continuum limit which use a sharp cut-off procedure. The resulting bubbles   can be classified into two   logarithmically divergent bubbles  of  the p-p channel at  zero momentum  pair and the particle-hole one at momentum $2\kf$, which leads to the most important contributions to the flow equations: 
\begin{equation}
\begin{split}
\label{Intengolgy}
\mathcal{L}_{\mathrm{C}} & = \pi \vf\,\La \dr_\La \int_{-\La_0}^{\La_0} \Theta(|\xi|-\La)  \, T \sum_{\omega_n}  G^0(k_\xi, \omega_n)G^0(-k_\xi, -\omega_n)\,\dd \xi,\cr
\mathcal{L}_{\mathrm{P}} & =  -  \pi\vf\, \La \dr_\La \int_{-\La_0}^{\La_0} \Theta(|\xi|-\La) \, T \sum_{\omega_n}  G^0(k_\xi -2k_{\rm F}, \omega_n)G^0(k_\xi, \omega_n)\,\dd \xi,\cr 
\mathcal{L}_{\mathrm{P'}} & =  -  \pi\vf\,\La \dr_\La \int_{-\La_0}^{\La_0} \Theta(|\xi|-\La) \, T \sum_{\omega_n}  G^0({k_\xi+2k_{\rm F}}, \omega_n)G^0(k_\xi, \omega_n)\,\dd \xi,
\end{split}
\end{equation}
where $ k_\xi = \arccos\big( - (\xi + \mu)/2  \big) $. The last contribution $\mathcal{L}_{\mathrm{P'}}$ is affected by the fact that the nesting relation is not perfect away from half-filling. As a consequence,  $\mathcal{L}_{\mathrm{P}}$ and $ \mathcal{L}_{\mathrm{P'}}$ differ in general, except at half-filling where $G=4\kf$ \cite{Seidel82}.   The second category comes from  non-divergent bubbles of p-p and p-h scattering channels (respectively noted $ \mathcal{L}_{\mathrm{C}'} $ and $ \mathcal{L}_{\mathrm{L}} $) when both particles belong to the same energy branch. These take the form 
\begin{align}
\label{IntengolgyL}
\mathcal{L}_{\mathrm{L},\mathrm{C}'} & = \mp\pi\vf\,\La \dr_\La\int_{-\La_0}^{\La_0}   \Theta(|\xi|-\La)
\, T \sum_{\omega_n}  G^0(k_\xi+0^+, \omega_n)G^0(k_\xi, \omega_n)\,\dd \xi
\end{align}
and only take finite values within the thermal shell $\Lambda\lesssim T$. 

To  determine the phase diagram, we have to derive further the flow equations for the three-leg vertices. This is done by adding to the effective action terms which couple the electronic field to the source field. One has to include terms associated to charge and  spin density waves, centered on sites or on bonds, and singlet/triplet superconductivity:
\begin{align}
\Gamma_{Z}[\phi, H, J] & = \sum_{a, a', b} \sum_{\sigma, \sigma'} \Big\{ Z^{\mathrm{ch}}_{ a  a', b} \bar{\phi}_{a \sigma} H^0_{ b   } \sigma^{0}_{\sigma \sigma'} \phi_{a' \sigma'}  +  Z^{\mathrm{sp}}_{ a  a', b} \bar{\phi}_{a \sigma} \vec{H}_{b   } \cdot \vec{\sigma}_{\sigma \sigma'} \phi_{a' \sigma'} \nonumber \\
 & \qquad \qquad \qquad + Z^{\mathrm{s}}_{a  a', b} \bar{\phi}_{a \sigma} J^0_{b } \pi^{0}_{\sigma \sigma'} \bar{\phi}_{a' \sigma'} + \mathrm{c.c.} \nonumber \\
 & \qquad \qquad \qquad + Z^{\mathrm{t}}_{ a  a', b} \bar{\phi}_{a \sigma} \vec{J}_{ b } \cdot \vec{\pi}_{\sigma \sigma'} \bar{\phi}_{a' \sigma'} + \mathrm{c.c.} \Big\} ,
\end{align}
where $ \sigma^0 $ is the $2 \times 2$ identity matrix,  $ \vec{\sigma} = (\sigma^1, \sigma^2, \sigma^3) $ is the vector containing the Pauli matrices, $ \pi^0 = - \ii \sigma^2 $ and $ \vec{\pi} = - \ii \sigma^2 \Vec{\sigma} $. Furthermore, the $Z$ vertices have the following expressions:
\begin{align}
Z^{\text{ch-s}}_{a' a ; c} & = Z^{\text{ch-s}}_k(q) \derr_{k - k' + q} \delta_{\omega_{n} - \omega_{n'} , 0} , \nonumber \\
Z^{\text{ch-b}}_{a' a ; c} & = Z^{\text{ch-b}}_k(q) \cos\big(  (k+q)/2 \big) \derr_{k - k' + q} \delta_{\omega_{n} - \omega_{n'} , 0}  ,   \\
Z^{\mathrm{s}/\mathrm{t}}_{a' a ; c} & = Z^{\mathrm{s}/\mathrm{t}}_{k}(q) \derr_{k + k' + q} \delta_{\omega_{n} + \omega_{n'}, 0} ,  \nonumber
\end{align}
with the correspondences given for static source fields $H$ et $J$:
\[
a \to (\omega_n, k), \quad a' \to (\omega_{n'}, k'), \quad b \to ( 0, q ).
\]
In the case of the $g$-ology model, we limit ourselves to the vertices $  Z^{\textsc{x}}_{\eta \kf }(q)  $ evaluated at $ q = \pm 2 \kf $ for the density waves, and at $ q = 0 $ for the singlet/triplet superconductivity. There are four density waves at $ 2\kf $ which correspond to site-centered charge- and spin-density wave (CDW, SDW), and bond-centered charge- and spin-density wave (BOW, BSDW):
\begin{equation}
\begin{split}
\label{zgologyd}
Z_{\mathrm{CDW}} & = Z^{\text{ch-s}}_{+\kf}(-2\kf)  =  Z^{\text{ch-s}}_{-\kf}(+2\kf), \quad Z'_{\mathrm{CDW}}  = Z^{\text{ch-s}}_{+\kf}(+2\kf)  =  Z^{\text{ch-s}}_{-\kf}(-2\kf), \cr
Z_{\mathrm{BOW}} & = Z^{\text{ch-b}}_{+\kf}(-2\kf) = Z^{\text{ch-b}}_{-\kf}(+2\kf), \quad Z'_{\mathrm{BOW}}  = Z^{\text{ch-b}}_{+\kf}(+2\kf) = Z^{\text{ch-b}}_{-\kf}(-2\kf), \cr
Z_{\mathrm{SDW}} & = Z^{\text{sp-s}}_{+\kf}(-2\kf) = Z^{\text{sp-s}}_{-\kf}(+2\kf), \quad Z'_{\mathrm{SDW}}  = Z^{\text{sp-s}}_{+\kf}(+2\kf) = Z^{\text{sp-s}}_{-\kf}(-2\kf), \cr
Z_{\mathrm{BSDW}} & = Z^{\text{sp-b}}_{+\kf}(-2\kf)  =  Z^{\text{sp-b}}_{-\kf}(+2\kf), \quad Z'_{\mathrm{BSDW}} = Z^{\text{sp-b}}_{+\kf}(+2\kf)  =  Z^{\text{sp-b}}_{-\kf}(-2\kf) .
\end{split}
\end{equation}
The vertices associated to singlet (SS) and triplet (TS) superconductivity  are given by 
\begin{equation}
\begin{split}
\label{zgologysc}
Z_{\mathrm{SS}} & = Z^{\mathrm{s}}_{+\kf}(0) + Z^{\mathrm{s}}_{-\kf}(0) , \cr
Z_{\mathrm{TS}} & = Z^{\mathrm{t}}_{+\kf}(0) - Z^{\mathrm{t}}_{-\kf}(0).
\end{split}
\end{equation}
The flow equations  associated to the  density-wave vertices are thus 
\begin{equation}
\begin{split}
\label{zdens}
\dfrac{\dd Z_{\mathrm{x}}}{\dd \ell} & = \dfrac{1}{2}   Z_{\mathrm{x}} \tilde{g}_{\mathrm{x}}, \cr
\tilde{g}_{\mathrm{CDW}} & = (g_2 - 2 g_1)\mathcal{L}_{\mathrm{P}} -  g_3\mathcal{L}_{\mathrm{P'}}, \quad
\tilde{g}_{\mathrm{SDW}}  = g_2\mathcal{L}_{\mathrm{P}} + g_3\mathcal{L}_{\mathrm{P'}}, \cr
\tilde{g}_{\mathrm{BOW}} & = (g_2 - 2 g_1)\mathcal{L}_{\mathrm{P}} +  g_3\mathcal{L}_{\mathrm{P'}}, \quad
\tilde{g}_{\mathrm{BSDW}}  = g_2\mathcal{L}_{\mathrm{P}} -  g_3\mathcal{L}_{\mathrm{P'}},
\end{split}
\end{equation}
while those for singlet and triplet superconductivity are
\begin{equation}
\begin{split}
\label{zsc}
\dfrac{\dd Z_{\mathrm{x}}}{\dd \ell} & = \dfrac{1}{2} Z_{\mathrm{x}} \tilde{g}_{\mathrm{x}}, \cr
\tilde{g}_{\mathrm{SS}} & = ( g_1 +  g_2)\mathcal{L}_{\mathrm{C}}, \quad
\tilde{g}_{\mathrm{TS}}  = (g_2 - g_1)\mathcal{L}_{\mathrm{C}},
\end{split}
\end{equation}
where $\ell$ is the so-called RG time, such that $ \La = \La_0 \ee^{-\ell} $. The initial conditions   are $Z_{\rm x}(\ell=0) =1$  for all channels x.

The  expression of the normalized susceptibility that stands for any of the above correlation channels     is given by 
\begin{equation}
    \chi_{\mathrm{x}}(\ell)
    \quad  = \int_0^\ell Z^2_{\mathrm{x}}(\ell') \, \big| \mathcal{L}_{\mathrm{x}}(\ell') \big| \dd\ell'
    \label{kix}
\end{equation}
 with  $\chi_{\rm x}(\ell=0)=0$  as initial condition.  
The  phase of the system is defined by the most singular susceptibility $\chi_\mathrm{x}$ and therefore the most singular $Z_{\mathrm{x}}$. We shall limit ourselves to the phases with the most important singularities. These  correspond to   $2\kf$ density-wave and superconducting phases at zero pairing momentum, which are governed by Eqs.~(\ref{zdens}) and (\ref{zsc}). The corresponding three-leg vertices can be expressed as $Z_\mathrm{x}(\ell)= \exp[ \frac{1}{2} \gamma_\mathrm{x}(\ell)] $, with a scale-dependent exponent  $\gamma_\mathrm{x}(\ell) = \int_0^\ell \tilde{g}_\mathrm{x}(\ell')\dd\ell'$.

\subsubsection{Half-filling}

It is useful in what follows to recall the main features of the one-loop flow equations of the continuum theory both at and away from half-filling.
 We first consider the case at half-filling and in the zero-temperature limit where $\mu =0$ ($n=1$) and  $\beta \to \infty$. This gives for the bubble intensities (\ref{Intengolgy}):
\begin{equation}
\begin{split}
            \mathcal{L}_{\mathrm{P},\mathrm{P'}} & = - \mathcal{L}_{\mathrm{C}} =  \tanh(\beta\La/2)\to 1,  \\
         \mathcal{L}_{\mathrm{L,C}'} &  = \mp 2\La\dr_\Lambda \FD(\La)\to 0,
    \label{Lloop}
\end{split}
\end{equation}
where $\FD$ is the Fermi distribution.
From (\ref{gologyg}) one recovers the well known $g$-ology flow equations at half-filling \cite{Dzyaloshinskii72,Kimura75,Solyom79}:
\begin{equation}
   \begin{split}
\label{gology}
\dfrac{\dd g_1}{\dd \ell} & = - g_1^2, \\
\dfrac{\dd g_2}{\dd \ell} & =  (g_3^2- g_1^2)/2, \\
\dfrac{\dd g_3}{\dd \ell} & =  g_3 ( 2 g_2 -g_1), \\
\dfrac{\dd g_4}{\dd \ell} & = 0.
\end{split} 
\end{equation}

If the coupling constants remain weak for all values of $\ell$ then the electron system evolves towards a Tomanaga-Luttinger (TL) liquid with $g_2$ and $g_4$ couplings only and gapless excitations. On the other hand, if the flow of either $g_1$ or $(2g_2-g_1,g_3)$  evolves towards a singularity at a critical $\ell_0$, the perturbative one-loop RG breaks down and we expect the formation of a gap $\varDelta =\La_0 \ee^{-\ell_0}$ in the spin ($g_1\to -\infty$) or charge ($2g_2-g_1 \to +\infty ,|g_3|\to +\infty$)\footnote{See also the footnote \ref{Gapcharge} below.} long-wavelength degrees of freedom.   

The flow of $g_1(\ell)$ associated to the spin degrees of freedom  is decoupled from those of $g_3(\ell)$ and  $2g_2(\ell)-g_1(\ell)$  linked to the charge ones. These  combine  to give  the scale invariant constant $C= g_3^2(\ell) - \big(2g_2(\ell)-g_1(\ell)\big)^2 $   \cite{Dzyaloshinskii72}. Thus for an initial attraction, $g_1<0$ ($U<2V$), the flow of $g_1(\ell)$ scales to  strong attractive coupling  with a singularity that develops at a finite $\ell_\sigma$, indicative of a spin gap $\varDelta_\sigma \sim \La_0 \ee^{-\ell_\sigma}$; whereas for an initial  repulsion $g_1>0$, $g_1(\ell)$ is marginally irrelevant and  spin  degrees of freedom remain gapless. For  the charge part, when $g_1-2g_2\ge |g_3|$, umklapp scattering  becomes marginally irrelevant, $2g_2(\ell)-g_1(\ell)$ then scales to a non-universal value and the charge-density sector remains gapless. By contrast, when  $g_1-2g_2< |g_3|$, the umklapp term is marginally relevant and the flow leads to a singularity in both $g_3(\ell)$ and $2g_2(\ell) -g_1(\ell) $ at $\ell_\rho$  implying a Mott gap $\varDelta_\rho \sim \La_0 \ee^{-\ell_\rho}= \La_0 \ee^{-1/\sqrt{|C|}}$ in the charge sector. Finally, at the one-loop level there are no logarithmic contributions to the flow of intra-branch forward scattering $g_4$, which remains scale invariant.

\begin{figure}
    \centering
    \includegraphics[scale=1.35]{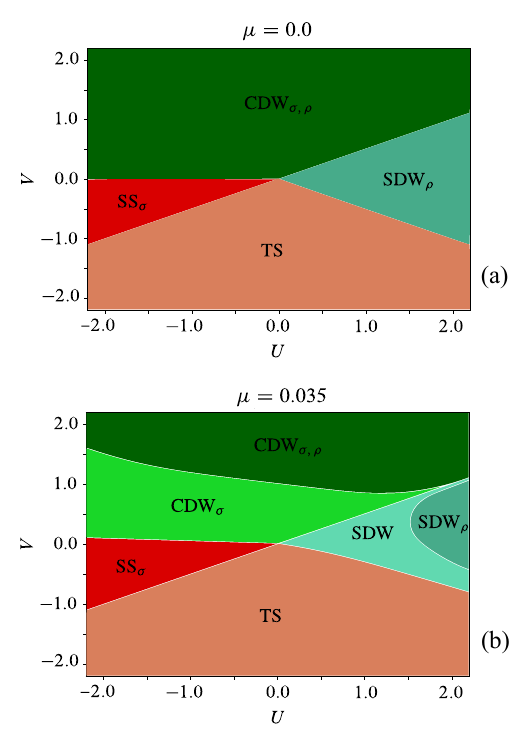}
   \caption{Weak-coupling phase diagram of the extended Hubbard model obtained from the $g$-ology model (continuum limit, linear spectrum and momentum-independent interactions) at  half-filling $\mu=0$ (a) and  small doping $\mu=0.035$ (b).  $U$ and $V$ are expressed in units of bare hopping $t$. The subscripts $\sigma$/$\rho$ of  a given phase indicate the presence of a gap in the spin/charge excitations (see also footnote \ref{Gapcharge}).}
    \label{Diag_g_ology}
\end{figure}

Regarding the phase diagram as a function of $U$ and $V$, when  ${U>\pm 2V}$, so that $g_1>0$ and $g_1-2g_2< |g_3|$,  the strongest singularity appears for $\chi_{\mathrm{SDW}}$, $\gamma_{\mathrm{SDW}}$ being the largest exponent of (\ref{zdens}), with  a SDW state having  gapless spin excitations  and a Mott gap.  For $V<\mp U/2$, so that $g_1>0$ and $g_1-2g_2>|g_3|$, (\ref{zdens}) yields   $\gamma_\mathrm{TS}$ as the largest  exponent and a dominant  susceptibility for TS with gapless excitations for both spin and charge. For $U/2 < V<0 $, which implies $g_1<0$ and $g_1-2g_2>|g_3|$,  it is in turn $\gamma_{\mathrm{ss}}$ to be the largest exponent in (\ref{zdens}) with a dominant singularity in the SS susceptibility with a spin gap. Finally when $ U/2 < V$  and $V >0 $, we have $g_1<0$  and $g_1-2g_2< |g_3|$ leading to a CDW phase, which  is gapped for both spin and charge excitations. Along the separatrix $U=2V$, $g_1=g_3=0$,  corresponding to gapless conditions of the TL model,  $\gamma_{\mathrm{SDW}}= \gamma_{\mathrm{CDW}}$ and $\chi_{\mathrm{CDW}}$ and $\chi_{\mathrm{SDW}}$ are equally singular at $U>0$, whereas at $U<0$, $\gamma_{\mathrm{SS}}= \gamma_{\mathrm{TS}}$, and  $\chi_{\mathrm{TS}}$ and $\chi_{\mathrm{SS}}$ are  equal. Finally, the symmetry line at $U<0$ and $ V=0$, with $g_1<0$ and $g_1-2g_2 >|g_3|$, leads to $ \gamma_{\mathrm{CDW}}=\gamma_{\mathrm{SS}} $ and coexisting  CDW and SS phases. The resulting well known   phase diagram  of the continuum theory is shown in Fig.~\ref{Diag_g_ology}~\cite{Kimura75}. It is worth noting that in the $g$-ology model,  $\chi_{\mathrm{BOW}}$ never  appears as the  dominant susceptibility, but only as the subdominant one in the SDW phase \cite{Kimura75}.  

\subsubsection{ Away from half-filling}

We now turn to the main results for finite values of $\mu$.  From  (\ref{gologyg}), the one-loop flow equations at finite doping  in the low-temperature limit can be put in the form
\begin{equation}
   \begin{split}
\label{gologymu}
&\dfrac{\dd g_1}{\dd \ell}  = - g_1^2, \\
&\dfrac{\dd} {\dd \ell}(2g_2-g_1)  =  g_3^2\mathcal{L}_{P'}, \\
&\dfrac{\dd g_3}{\dd \ell}  =  (1 +\mathcal{L}_{P'})  g_3 (2 g_2 -g_1)/2, \\
&\dfrac{\dd g_4}{\dd \ell} = 0.
\end{split} 
\end{equation}
These equations correspond to the former results of Seidel {\it et al.} \cite{Seidel82,Firsov85} and are consistent with those of the bosonization approach in the weak-coupling limit \cite{Giamarchi04,Giamarchi91,Montambaux86}. We illustrate this situation at a finite but small doping  $\mu=0.035$. 

The flow  of  $g_1$, tied to the spin degrees of freedom, keeps the same form as before, namely $g_1(\ell)= g_1(1+ g_1\ell)^{-1} $, indicating a spin gap $\varDelta_\sigma \sim \La_0\ee^{-1/|g_1|}$ when $g_1<0$ i.e. in the region above the separatrix $V=U/(2-\mu^2)$ whose slope increases with  $\mu$, as shown in Fig.~\ref{Diag_g_ology}-(b). Concerning the charge degrees of freedom, a finite $\mu$ affects the flows of  $2g_2(\ell) -g_1(\ell)$ and $g_3(\ell)$ due to  the  suppression  of the logarithmic singularity of the particle-hole loop ${\cal L}_{P'}$ when $\La(\ell)< \mu$. Thus at sufficiently small couplings, the flow of $2g_2(\ell) -g_1(\ell)$ is no longer singular so that  no charge gap  is   possible. This introduces   gapless regions for the charge sector (Fig.~\ref{Diag_g_ology}-(b)) corresponding to either CDW or SDW phases.  By cranking up $\mu$, the charge-gapped regions persist\footnote{The   gap $\varDelta_\rho$ that persists in the RG flow for some interval of $\mu\ne 0\, (n\ne 1)$ refers to the energy distance between the upper and lower Hubbard like sub-bands, which does not contract  immediately to zero by doping away from half-filling.  This finite excitation energy should not be confused with the Mott insulating gap which immediately closes at the metal-insulator transition when  $n \to 1^\pm $. It is only at half-filling that this energy distance coincides with the insulating gap. \label{Gapcharge}  } but shrink in size being pushed to higher couplings. In the gapless-charge domains,  umklapp scattering reduces to a simple renormalization of the combination $2g_2-g_1$, which becomes scale invariant  in weak coupling [see Eq.~(\ref{gologymu})]. In the  $g_1>0$ part of  Fig.~\ref{Diag_g_ology}-(b), that is for    $V<U/(2-\mu^2)$, the detrimental effect of doping on  umklapp is also apparent for the gapless region where the most  important power-law singularity in $\chi_{\rm TS}$ gains in importance against SDW. A similar effect takes place at $g_1<0$ where the  SS region, in which umklapp scattering is an irrelevant coupling, gains in importance against CDW when $\mu$ increases. 

We shall examine  next to what extent taking into account  lattice effects of the EHM model can alter these  results.

\subsection{Lattice effects and low-energy limit}

Lattice effects are twofold. First, they are present in the one-body term of the Hamiltonian  through the  inter-site hopping of electrons. This leads to the tight-binding spectrum of Fig.~\ref{TightBinding} showing  the growth of its curvature as energy moves away  from the Fermi level and becoming particle-hole  asymmetric  away from half-filling.  Second, they appear in the coupling constants that are spatially non-local. This is the case of the nearest-neighbor interaction $V$  which introduces a dependence on wave vectors in momentum space.  

Both effects are  linked  since  the  momentum dependence  of interactions generates an additional curvature of the spectrum through one-particle self-energy corrections. At the one-loop level, these come from  Hartree-Fock contributions to the flow. However, as shown in appendix \ref{Fermi_velocity} those corrections are small in weak coupling and will therefore be ignored in the following.

We now turn to the effects of the lattice on coupling constants.  When defined on the two Fermi points, like the $g$'s  of the $g$-ology continuum theory considered above,  they are known to be marginal. The momentum dependence of the coupling constants is irrelevant in the RG sense but can have both qualitative and quantitative effects on the phase diagram.

In order to classify the coupling constants, it is advantageous to consider the energy variables rather than the momenta \cite{Menard11,Dumoulin96}. This is done using the   dispersion $ \xi(k) $ as  measured with respect to the Fermi level. In fact, there is a one-to-one correspondence between momenta $k$ on the one hand, and $ (\xi(k), \eta) $ on the other hand, where $ \eta = \sgn k$ correspond  to the branch index for positive $k$ $(\eta=+)$ and negative $k$ $(\eta=-)$.  The idea is therefore to introduce a systematic expansion of the  vertices and loops in power of the   $\xi$  variables  tied to the independent momenta $k$. For the couplings, one has
\begin{equation}
    \begin{split}
g_{k_1, \, k_2, \,  k'_1} & = g^{\vec \eta}(\vec \xi) = \sum_{n_i = 0}^{\infty} \dfrac{ \xi_1^{n_1} \xi_2^{n_2} \xi_{1'}^{n'_1} }{n_1! n_2 ! n'_1!} g^{\vec \eta}_{\vec n} , \label{XiDev}  \\
\vec x & = (x_1, x_2, x_{1'}), \quad x = \xi, \eta, n.
\end{split}
\end{equation}
 where the coefficients  $g^{\vec \eta}_{\vec n}$   now stand   for the set of marginal and irrelevant interactions of the model.

We now derive the general form of the flow equations when the $ \xi $ expansion is made explicit. One first makes the change of variables $ k \rightarrow (\eta, \xi) $, and writes
\begin{equation}
    \dfrac{\dd g^{\vec \eta}_{\vec n}}{\dd \ell} = - \La \dr_\La  g^{\vec \eta}_{\vec n} =  \sum_{\mathrm{x}} D^{\vec \eta}_{\mathrm{x}}(\vec \xi),
\end{equation}
where the sum runs over all Feynman graphs of Fig.~\ref{dgDiag}, that is to say $ \mathrm{x} \in \{\mathrm{p}, \mathrm{ph}1, \mathrm{ph}2, \mathrm{ph}3   \} $. Furthermore, each diagram can be written in the form
\begin{equation}
     D^{\vec \eta}_{\mathrm{x}}(\vec \xi) = \sum_{p} \, \mathcal{L}^{\vec \eta}_{\mathrm{x}} (p, \vec \xi) \, \gamma^{\vec \eta}_{\mathrm{x}1} (p, \vec \xi) \,  \gamma^{\vec \eta}_{\mathrm{x}2} (p, \vec \xi),
\end{equation}
where $ \mathcal{L}^{\vec \eta}_{\mathrm{x}} (p, \vec \xi) $ is a bubble of the scattering channel $\mathrm{x}$, while the $ \gamma^{\vec \eta}_{\mathrm{x}i} (p, \vec \xi) $'s are combinations of the coupling constants. For example, in the case $ \mathrm{x} = \mathrm{pp} $, one gets from Eqs.~(\ref{Dee}) of  Appendix:
\begin{align*}
    \mathcal{L}^{\vec \eta}_{\mathrm{pp}} (p, \vec \xi) & = \mathcal{L}^{\mathrm{pp}}_{p, -p+k_1+k_2}, \\
    \gamma^{\vec \eta}_{\mathrm{x}1} (p, \vec \xi) & = g_{k_2, k_1, -p+k_1 +k_2}, \\
    \gamma^{\vec \eta}_{\mathrm{x}2} (p, \vec \xi) & =  g_{p, -p+k_1 + k_2, k'_1}.
\end{align*}
The corresponding expressions for the other channels $\mathrm{x}=\mathrm{ph1},  \mathrm{ph2}$ and $ \mathrm{ph3}$ are given in Eqs.~(\ref{Deh1}), (\ref{Deh2}) and (\ref{Deh3}), respectively. It is then possible to make use of the expansion given in \eqref{XiDev} for the couplings and a similar one  for the bubbles. Once this is done, the flow equations are written as 
\begin{equation}
       \sum_{n_i = 0}^{\infty} \dfrac{ \xi_1^{n_1} \xi_2^{n_2} \xi_{1'}^{n'_1} }{n_1! n_2 ! n'_1!}\dfrac{\dd  g^{\vec \eta}_{\vec n}}{\dd \ell} 
       = \sum_{n_i = 0}^{\infty} \sum_{m_{1,i}=0}^\infty\sum_{m_{2,i}=0}^\infty\dfrac{ \xi_1^{n_1} \xi_2^{n_2} \xi_{1'}^{n'_1} }{n_1! n_2 ! n'_1!}   \mathcal{L}_{\vec n, \vec m_1, \vec m_2 }^{\vec \eta, \vec \eta_1, \vec \eta_2}  g^{\vec \eta_1}_{\vec m_1}  g^{\vec \eta_2}_{\vec m_2} .
       \label{FlowXi}
\end{equation}
Now it is useful to express the flow equations in a dimensionless form. Let us introduce the dimensionless quantities  $\tilde g^{\vec \eta}_\La( \tilde{\vec \xi})$, where $ \tilde{\vec \xi}= \vec \xi/\La$. The natural unit is  the cut-off $ \La $, 
\begin{align}
    g^{\vec \eta}_\La(\vec \xi) = \La^{[g]} \tilde g^{\vec \eta}_\La( \vec \xi /\La \, ) 
    \iff \tilde g^{\vec \eta}_\La( \tilde{\vec \xi}   \, ) = \La^{-[g]} g^{\vec \eta}_\La(\La  \tilde{\vec \xi}\, ) .
    \label{getadef}
\end{align}
 In this expression,  $ [g] $ denotes the engineering dimension of the coupling constant $g$. For two-body interactions in one dimension,
 energy-independent coupling constants are dimensionless, $ [g] = 0 $. From the expansion \eqref{XiDev}, it is straightforward to determine the dimension of a generic coupling constant:
\begin{equation}
    g^{\vec \eta}_{\vec n} =  \La^{-|\vec n |} \tilde g^{\vec \eta}_{\vec n} ,
\end{equation} 
where the notation $ | \vec n | = n_1 + n_2 + n_{1'} $ has been introduced.
The dimensionless flow equations for the coupling constants are then obtained by a simple identification from \eqref{FlowXi}:
\begin{equation}
       \dfrac{\dd \tilde g^{\vec \eta}_{\vec n}}{\dd \ell} =  -\La \dr_\La \tilde g^{\vec \eta}_{\vec n} = - | \vec n | \tilde g^{\vec \eta}_{\vec n}  - \sum_{n_{1,i}=0}^\infty\sum_{n_{2,i}=0}^\infty \tilde{ \mathcal{L}}_{\vec n, \vec n_1, \vec n_2 }^{\vec \eta, \vec \eta_1, \vec \eta_2}  \tilde{g}^{\vec \eta_1}_{\vec n_1}  \tilde{g}^{\vec \eta_2}_{\vec n_2},
       \label{gflowlat}
\end{equation}
with $  \tilde{ \mathcal{L}}_{\vec n, \vec n_1, \vec n_2 }^{\vec \eta, \vec \eta_1, \vec \eta_2} = \La^{|\vec n | - |\vec n_1 | - | \vec n_2 |} \mathcal{L}_{\vec n, \vec n_1, \vec n_2 }^{\vec \eta, \vec \eta_1, \vec \eta_2} $.
As a consequence, the expansion in $\xi$ classifies the coupling constants by order of irrelevance from the value of $|\vec{n}| $. In practice, we will restrict ourselves  to  quadratic order, i.e. $|\vec{n}|\leq 2$. Let us also note that it would be possible to expand the vertices in power of the Matsubara frequencies\cite{Markhof18,Yirga21}. It follows from dimensional analysis that the terms containing non-zero powers of the Matsubara frequencies are irrelevant. Since these terms are not present in the initial action (the interactions are not retarded), they can only be generated by the flow and are thus expected to remain negligible.

The different sets of interactions and  their initial conditions can be  expressed in terms of the coupling constants  of the original \ehm ; see Eq.~(\ref{gInit}). 
The expansion of the cosine in terms of the variables $ (\xi, \eta) $ gives, up to second order in $\xi$,
\begin{align}
\cos(k_1 -k'_{1} ) ={}& \eta_1 \eta_{1'} + (1-\eta_1 \eta_{1'}) \frac{\mu^2}{4}  
+  \dfrac{\mu }{4}(1-\eta_1 \eta_{1'}) ( \xi_1 + \xi_{1'}  ) \nonumber \\
& \ \ - \eta_1 \eta_{1'} \left(  \dfrac{1}{8}  -  \dfrac{\mu^2}{32(1-\mu^2/4)}  \right)  ( \xi_1^2 +  \xi_{1'}^2 )  \nonumber \\ & 
+ \left( \dfrac{1}{4} +  \eta_1 \eta_{1'} \dfrac{\mu^2}{16(1-\mu^2/4)} \right) \xi_1 \xi_{1'} + ... .
\label{expand}
\end{align}
Hence we obtain the following initialization conditions for the coupling constants  introduced in (\ref{XiDev}). For marginal interactions ${\mathcal{O}(\xi^0)}$  {($\vec{n}=0$)}, one has
\begin{equation}
\begin{split}
  g^{+\eta, -\eta, -\eta}_{0, 0, 0}  = g_{1} , \qquad &  g^{+\eta, -\eta, +\eta}_{0, 0, 0}  = g_{2},\cr
  g^{+\eta, +\eta, -\eta}_{0, 0, 0}  = g_{3} , \qquad &
  g^{+\eta, +\eta, +\eta}_{0, 0, 0}  = g_{4},
\end{split}
\end{equation}
and the initial values coincide with those of the continuum theory in (\ref{gini}).

From (\ref{expand}) and   at ${\mathcal{O}(\xi)}$  ($|\vec{n}|=1$),  the set of irrelevant interactions labeled in terms of backward, forward and umklapp scattering amplitudes, together with their  initial filling-dependent values, reads:
\begin{equation}
    \begin{split}
\label{gxi0}
 & g^{+\eta, -\eta, -\eta}_{1, 0, 0} = \frac{V\mu}{\pii \vf}, \quad  g^{+\eta, -\eta, -\eta}_{0, 0, 1}  = \frac{V\mu}{\pii \vf} , \quad 
  g^{+\eta, -\eta, +\eta}_{1, 0, 0} =0, \quad  g^{+\eta, -\eta, +\eta}_{0, 0, 1}  =0, \\ &
  g^{+\eta, +\eta, -\eta}_{1, 0, 0}  = \frac{V\mu}{\pii \vf}, \quad  g^{+\eta, +\eta, -\eta}_{0, 0, 1}  = \frac{V\mu}{\pii \vf}, \quad g^{+\eta, +\eta, +\eta}_{1, 0, 0}  =0, \quad g^{+\eta, +\eta, +\eta}_{0, 0, 1} =0.
\end{split}
\end{equation}

 Likewise, the set of irrelevant couplings  at ${\mathcal{O}(\xi^2)}$ ($|\vec{n}|=2$) and  their  initial values can be put in the form
 \begin{equation}
\begin{split}
\label{gxi20}
    g^{+\eta, -\eta, -\eta}_{2, 0, 0}  =  \dfrac{V}{\pii \vf}  \left( \dfrac{1}{2} - \dfrac{\mu^2}{8(1-\mu^2/4)} \right),  & \qquad g^{+\eta, -\eta, -\eta}_{0, 0, 2}  = \dfrac{V}{\pii \vf}  \left( \dfrac{1}{2} - \dfrac{\mu^2}{8(1-\mu^2/4)} \right) , \cr
    g^{+\eta, -\eta, -\eta}_{1, 0, 1} = \dfrac{V}{\pii \vf}  \left( \dfrac{1}{2} - \dfrac{\mu^2}{8(1-\mu^2/4)} \right), & \qquad  g^{+\eta, -\eta, +\eta}_{2, 0, 0}  = -\dfrac{V}{\pii \vf}  \left( \dfrac{1}{2} - \dfrac{\mu^2}{8(1-\mu^2/4)} \right), \cr 
    g^{+\eta, -\eta, +\eta}_{0, 0, 2}  = -\dfrac{V}{\pii \vf}  \left( \dfrac{1}{2} - \dfrac{\mu^2}{8(1-\mu^2/4)} \right),  & \qquad 
    g^{+\eta, -\eta, +\eta}_{1, 0, 1}  = \dfrac{V}{\pii \vf}  \left( \dfrac{1}{2} + \dfrac{\mu^2}{8(1-\mu^2/4)} \right), \cr
    g^{+\eta, +\eta, -\eta}_{2, 0, 0}  = \dfrac{V}{\pii \vf}  \left( \dfrac{1}{2} - \dfrac{\mu^2}{8(1-\mu^2/4)} \right), & \qquad
    g^{+\eta, +\eta, -\eta}_{0, 0, 2}  = \dfrac{V}{\pii \vf}  \left( \dfrac{1}{2} - \dfrac{\mu^2}{8(1-\mu^2/4)} \right), \cr
    g^{+\eta, +\eta, -\eta}_{1, 0, 1}  = \dfrac{V}{\pii \vf}  \left( \dfrac{1}{2} - \dfrac{\mu^2}{8(1-\mu^2/4)} \right), & \qquad 
    g^{+\eta, +\eta, +\eta}_{2, 0, 0}  = -\dfrac{V}{\pii \vf}  \left( \dfrac{1}{2} - \dfrac{\mu^2}{8(1-\mu^2/4)} \right),  \cr
    g^{+\eta, +\eta, +\eta}_{0, 0, 2}  = -\dfrac{V}{\pii \vf}  \left( \dfrac{1}{2} - \dfrac{\mu^2}{8(1-\mu^2/4)} \right), & \qquad  
    g^{+\eta, +\eta, +\eta}_{1, 0, 1}  = \dfrac{V}{\pii \vf}  \left( \dfrac{1}{2} + \dfrac{\mu^2}{8(1-\mu^2/4)} \right). 
\end{split}
 \end{equation}

The same expansion procedure can in principle be applied  to  the vertex parts of the response functions:
\begin{align}
Z^{\textsc{x}}_{k}(q)  & =  \sum_{n=0}^{\infty} \dfrac{ \xi^n }{n!} Z^{\textsc{x}}_{\eta, \, n}(q) .\label{XiDevZ}
\end{align}
However, at variance with the coupling constants, the irrelevant contributions to all $Z_{\mathrm{x}}$ are zero at $\ell=0$, so that their effect on the flow will be negligible.  In the following, we shall therefore proceed to the evaluation of $Z^{\textsc{x}}_{k}(q)$ in the lowest or marginal  order by retaining only  $  Z^{\textsc{x}}_{\eta, \, n=0}(q)  $. Higher-order corrections are not expected to bring any qualitative modifications to the phase diagram.  Thus for the site- and bond-density-wave channels at $q=\pm 2\kf$, the  flow equations are respectively
\begin{equation}
\begin{split}
\label{zphflowa}
\La\dr_\La Z'_{ \mathrm{CDW} } & = {\dfrac{1}{2}} \mathcal{L}_{\mathrm{P}'}(g_2 - 2 g_1)Z'_{ \mathrm{CDW} } -  {\dfrac{1}{2}}\mathcal{L}_{\mathrm{P}} g_3Z_{ \mathrm{SDW} },  \cr
\La\dr_\La Z_{ \mathrm{CDW} } & = {\dfrac{1}{2}} \mathcal{L}_{\mathrm{P}} (g_2 - 2 g_1) Z_{ \mathrm{CDW} }-  {\dfrac{1}{2}}\mathcal{L}_{\mathrm{P}'}  g_3 Z'_{ \mathrm{CDW} },  \cr
\La\dr_\La Z'_{ \mathrm{SDW} }  & = {\dfrac{1}{2}} \mathcal{L}_{\mathrm{P}'} g_2 Z'_{ \mathrm{SDW} } + {\dfrac{1}{2}}\mathcal{L}_{\mathrm{P}}  g_3 Z_{ \mathrm{SDW} },  \cr
\La\dr_\La Z_{ \mathrm{SDW} }  & = {\dfrac{1}{2}}\mathcal{L}_{\mathrm{P}} g_2 Z_{ \mathrm{SDW} } + {\dfrac{1}{2}} \mathcal{L}_{\mathrm{P}'}  g_3 Z'_{ \mathrm{SDW} } ,  
\end{split}
\end{equation}
and 
\begin{equation}
\begin{split}
\label{zphflowb}
\La\dr_\La Z'_{ \mathrm{BOW} } & =  {\dfrac{1}{2}} \mathcal{L}_{\mathrm{P}'}   (g_2 - 2 g_1) Z'_{ \mathrm{BOW} } -  {\dfrac{1}{2}}\dfrac{\mathcal{L}_{\mathrm{P}}  }{ \cos(2\kf)  }   g_3 Z_{ \mathrm{BOW} },  \cr
\La\dr_\La Z_{ \mathrm{BOW} } & = {\dfrac{1}{2}} \mathcal{L}_{\mathrm{P}}  (g_2 - 2 g_1) Z_{ \mathrm{BOW} } -   {\dfrac{1}{2}}\mathcal{L}_{\mathrm{P}'}\cos(2\kf)  g_3 Z'_{ \mathrm{BOW} },  \cr
\La\dr_\La Z'_{ \mathrm{BSDW} }   &= {\dfrac{1}{2}} \mathcal{L}_{\mathrm{P}'}      g_2  Z'_{ \mathrm{BSDW} }   +   {\dfrac{1}{2}}\dfrac{ \mathcal{L}_{\mathrm{P}} }{  \cos(2\kf)  }    g_3  Z_{ \mathrm{BSDW} },  \cr
\La\dr_\La Z_{ \mathrm{BSDW} }   &= {\dfrac{1}{2}} \mathcal{L}_{\mathrm{P}}     g_2  Z_{ \mathrm{BSDW} }     +   {\dfrac{1}{2}}\mathcal{L}_{\mathrm{P}'} \cos(2\kf)   g_3  Z'_{ \mathrm{BSDW} }. 
\end{split}
\end{equation}
In the superconducting channel at zero pair momentum, one has 
\begin{equation}
    \begin{split}
\label{zsflow}
\La\dr_\La Z_{\mathrm{SS}} & = {\dfrac{1}{2}} \mathcal{L}_{\mathrm{C}} ( g_1 + g_2  )Z_{\mathrm{SS}} ,  \cr 
\La\dr_\La Z_{\mathrm{TS}}  &= - {\dfrac{1}{2}}\mathcal{L}_{\mathrm{C}} (g_1 - g_2 )Z_{\mathrm{TS}} .
\end{split}
\end{equation}

All the $Z_{\rm x}$ equations are bound to the initial  conditions $Z_{\rm x}(\ell=0)=1$. From these  the normalized susceptibilities  $\chi_{\rm x}$ in the channel x  can be obtained from the definition   (\ref{kix}) with initial condition $\chi_{\rm x}(\ell=0)=0$.
The main differences with respect to the $g$-ology model are in the bubbles and in the fact that the marginal coupling constants are influenced by the irrelevant ones.

We close this subsection  by considering the uniform $q\to 0$ response for charge and spin densities corresponding to  the uniform charge compressibility $(\chi_\rho)$ and spin susceptibility  $(\chi_\sigma)$ whose divergences signal the occurrence of phase separation and ferromagnetism. In the framework of the fRG, these susceptibilities can be easily computed using the fact that the degrees of freedom contributing to the p-p and $2k_F$ p-h fluctuations come from non thermal energies $|\xi|\gtrsim T$. These are separated  from those contributing to the $q\to 0$ response functions which rather correspond to the thermal width $|\xi|\lesssim T$. We then integrate first the flow equations (\ref{gflowlat}) considering  the  Cooper and Peierls channels alone with $\La$ running between $\La_0$ and $T$. The renormalized marginal coupling constants thus obtained  at the energy scale $T$ are then used to compute the uniform susceptibilities. Instead of integrating the flow with $\La$ running between $T$ and 0, one can simply use an RPA, which is known to be exact for a linear spectrum (and equivalent to bosonization) once fluctuations due to back scattering and umklapp processes have been integrated out. For a sufficiently small temperature, the results effectively correspond to the $T=0$ limit.

Explicitly one considers the uniform three-leg vertices which  obey the flow equation 
\begin{align}   
\dfrac{\dd Z_{\mathrm{x}}}{\dd \ell} & = \dfrac{1 }{4} \mathcal{L}_{\mathrm{L}} Z_{\mathrm{x}} g_{\mathrm{x}},
\end{align}
where
\begin{equation}
   \begin{split}
 g_{\mathrm{x}=\rho}(\ell) & = g_1(\ell) - 2g_2(\ell) -  g_4(\ell) ,\\
g_{\mathrm{x}=\sigma} & = g_1(\ell) + g_4(\ell). 
\end{split} 
\end{equation}
By inserting the one-loop RPA contributions to $g_{\mathrm{x}}(\ell)= g_{\mathrm{x}}^*/(1- \frac{1}{2}g_{\mathrm{x}}^*\chi_0(\ell))$ in the Landau scattering channel, one gets $Z_{\rho,\sigma}(\ell) = [1- \frac{1}{2}g_{\rho,\sigma}^*\chi^0(\ell)]^{-1}$ where $\chi^0(\ell)  =\frac{1}{2} \int_0^\ell \mathcal{L}_{\mathrm{L}}(\ell') \dd \ell'$ which according to (\ref{Lloop}) gives a non-zero contribution when the integration  $\ell'$ enters in the thermal energy interval $(\La(\ell')\lesssim T)$. Here $g_\rho^*= g_1^* - 2g_2^* -  g_4$  and $g_\sigma^*= g_1^* +  g_4$ and the starred  $g_{1,2}^*$ couplings are the renormalized values obtained from (\ref{gflowlat}) down to the edge of the thermal interval $\La\sim T$.  Following the definition of susceptibilities, 
\begin{equation}
 \chi_{\rho,\sigma}(\ell)
    \quad = \int_0^\ell Z^2_{\rho,\sigma}(\ell') \,\mathcal{L}_{\mathrm{L}}(\ell') \dd\ell',
    \label{kix}   
\end{equation}
one gets the expression 
\begin{equation}
    \chi_{\rho,\sigma}(T\to 0) = \frac{2}{1- \frac{1}{2} g_{\rho,\sigma}^*} 
    \label{Runif}
\end{equation}
for the normalized uniform compressibility and spin susceptibility in the zero temperature limit. These coincides with the expressions derived by the functional integral  method\cite{Nelisse99}.

\newpage
\section{Lattice model: results and discussion}

In this section, we will discuss the consequences of lattice effects  coming from the non-linearity of the spectrum and the momentum dependence of interactions in the determination of quantum phases of the EHM  as a function of filling. All calculations are carried out at the arbitrary chosen  temperature $T=10^{-7}$ which regularizes the Fermi distribution functions while being consistent with the zero temperature limit. The calculations are also limited to the weak-coupling sector.

\subsection{Half-filled case}
Before considering non-zero values of the chemical potential, let us examine as a benchmark of  our method the extensively studied half-filling case. The tight-binding spectrum  at $\mu=0$ shows a non-vanishing curvature as one moves away from the Fermi points $\pm k_{\mathrm{F}}$. On the boundaries  and at the center of the Brillouin zone, the spectrum displays a vanishing slope, which causes the appearance of a van Hove singularity. At half-filling    the progressive integration of degrees of freedom is then symmetric with respect to occupied and empty states.

\begin{figure}
    \centering
    \includegraphics[scale=0.55]{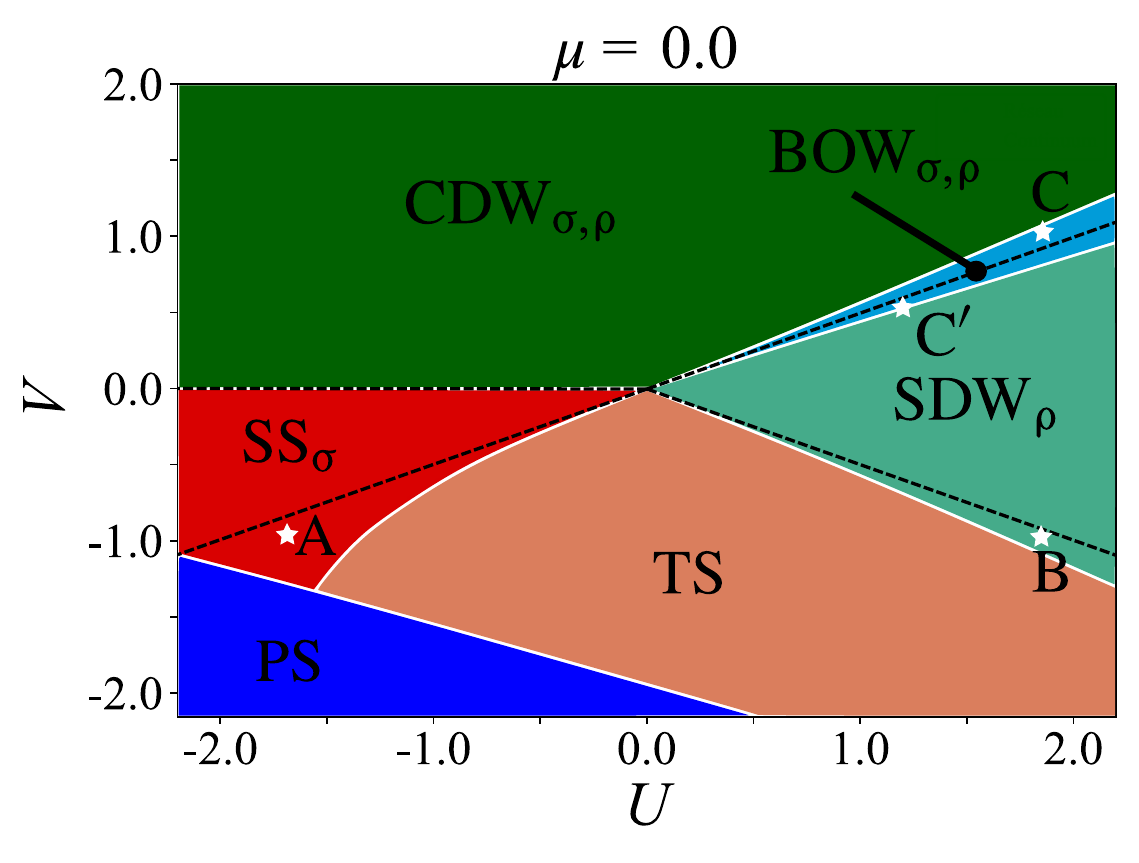}
    \caption{Phase diagram of the EHM at half filling. The points A, B, C and C' are discussed in the text. The dashed lines correspond to the phase boundaries of the continuum limit of the model shown in Fig.~\ref{Diag_g_ology}-(a).}
    \label{HalfFillingDiagA}
\end{figure}

 From the integration of Eqs.~(\ref{gflowlat})  and (\ref{zphflowa}-\ref{zsflow}), and by using the intial conditions (\ref{gini}) and (\ref{gxi0}-\ref{gxi20}) for the couplings, one obtains the half-filling EHM phase diagram shown in Fig.~\ref{HalfFillingDiagA}. Among the most striking modifications with respect to the continuum $g$-ology phase diagram  of Fig.~\ref{Diag_g_ology}-(a),  we first note the phases  located in the vicinity of the line $U = 2V$.  Recall  that in the $g$-ology framework,  both $g_1$ and $g_3$ vanish along that line at half-filling,  which leads to the conditions of the TL  model; crossing the line  then corresponds to a change of sign of  the  $g_1$ and $g_3$  coupling constants (see Eq. (\ref{gini})).
 
\begin{figure}
    \centering
    \includegraphics[scale=0.5]{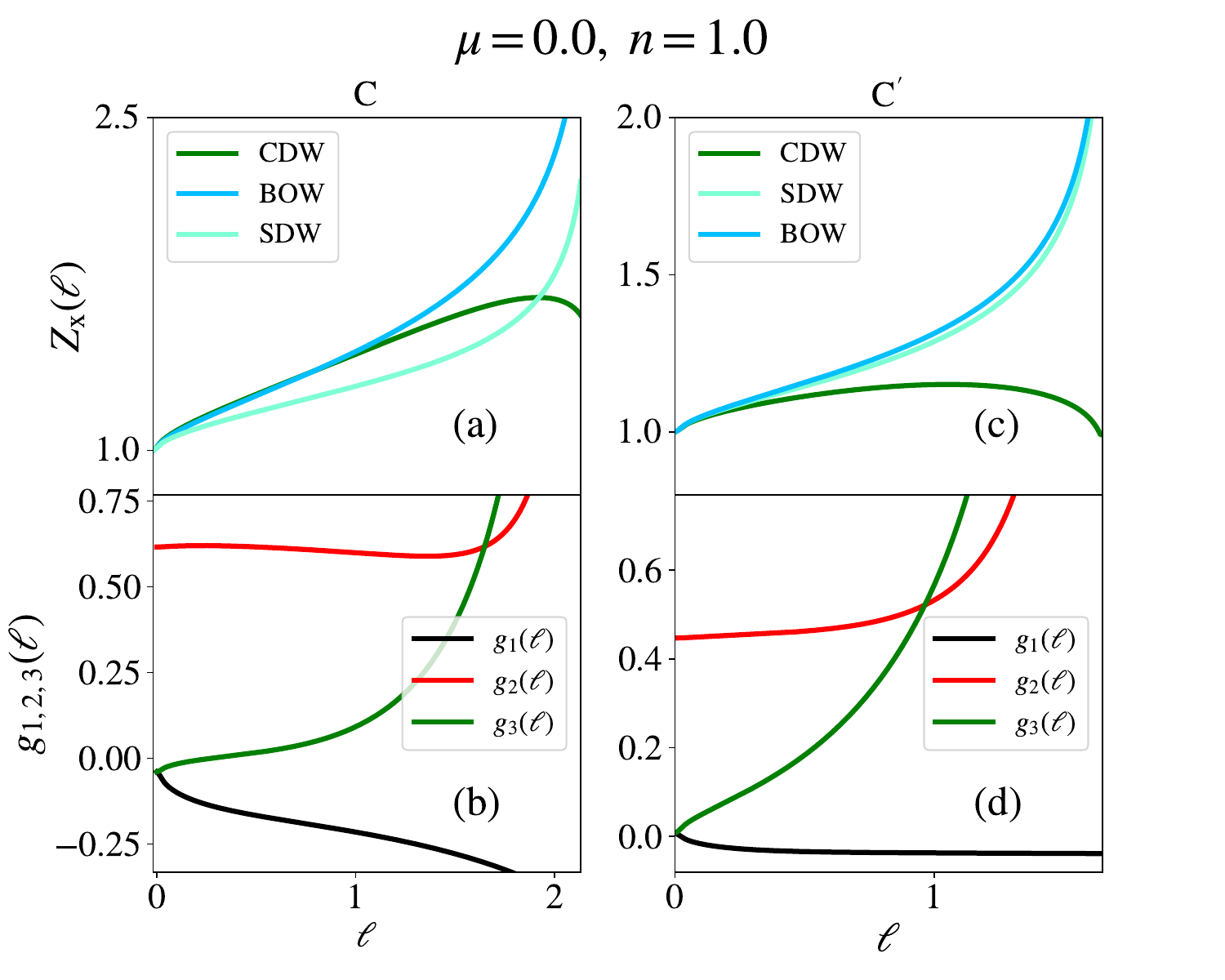}
    \caption{Flow of the three-leg vertices $Z_{\mathrm{x}}$ of the susceptibilities [(a),(c)] and coupling constants [(b),(d)] for points C' and C of the phase diagram in  Fig.~\ref{HalfFillingDiagA}, near the $U=2V>0$ line at half-filling. C:(1.81, 1.03), C':(1.43, 0.69).}
    \label{FlowC1C2}
\end{figure}

Along the line $U=2V>0$ in the repulsive part of the diagram, the gapless regime of the TL model  with equally singular SDW and CDW suceptibilities is made unstable by the presence of irrelevant couplings. Thus below but close to the line $U=2V$, at the point C' in the phase diagram of  Fig.~\ref{HalfFillingDiagA}, $g_3$ evolves to positive values and then  becomes relevant  together with the combination $2g_2-g_1$. Both diverge at some critical $\ell_\rho$, indicative of a charge (Mott) gap. The fate of $g_1$ is of particular interest since though repulsive initially,  it evolves toward negative values and its flow  ultimately separates from those  of $g_3$ and $2g_2-g_1$ at sufficiently large $\ell$ where the influence of irrelevant terms in (\ref{gflowlat}) at $|n|\ne 0$ becomes vanishinghly small and can be ignored above some arbitrary  value $\ell^*$ or equivalently below an effective cutoff energy $\varLambda^*= \varLambda e^{-\ell^*}$.  One finally recovers the flow of the continuum-limit theory [Eq.~(\ref{gology})], implying
\begin{equation}
g_1(\ell) = \dfrac{g_1(\ell^*)}{1+g_1(\ell^*)(\ell-\ell^*)} \qquad  (\ell\geq \ell^*),
\label{g1star}
\end{equation} 
where $g_1(\ell^*)<0$. Typically, we have $ |g_1(\ell^*)| \ll 1 $, so that the singularity of (\ref{g1star}) will invariably lead to a finite, though very small, gap  $\varDelta_{\sigma}\sim \varLambda^*\ee^{-1/|g_1(\ell^*)|}$ in the spin sector.   Slightly above the $U=2V$ line  at the point C in the phase diagram, both $g_1$ and $g_3$ are initially attractive. While $g_1$ remains attractive and evolves to strong coupling with the formation of a spin gap $\varDelta_{\sigma}$, which is much stronger in comparison  to C', the coupling $g_3$, though initially attractive, changes sign and becomes repulsive at the beginning of the flow due to its coupling to irrelevant terms. According to Fig.~\ref{FlowC1C2}-(b), the flows of $g_3$ and $2g_2-g_1$ then evolve to strong coupling and lead to the formation of a charge gap $\varDelta_\rho$.  

The consequence of effective repulsive $g_3$ and attractive $g_1$ couplings on the nature of correlations  is significant. On  the $U=2V$ line, instead of the coexistence of gapless CDW and SDW phases predicted by the TL model,  a spin and charge gapped BOW phase emerges. 
According to Figs.~\ref{FlowC1C2}-(a),(c), the gapped BOW state extends on either side of the line defining a fan shape  region where  it dominates over SDW and CDW phases.  These   findings confirm previous RG results \cite{Tsuchiizu02,Tam06,Menard11,Xiang19}, and are consistent with those of numerical   simulations in the weak-coupling region of the phase diagram  \cite{Nakamura00,Sengupta02,Ejima07}.
\begin{figure}
    \centering
    \includegraphics[scale=0.5]{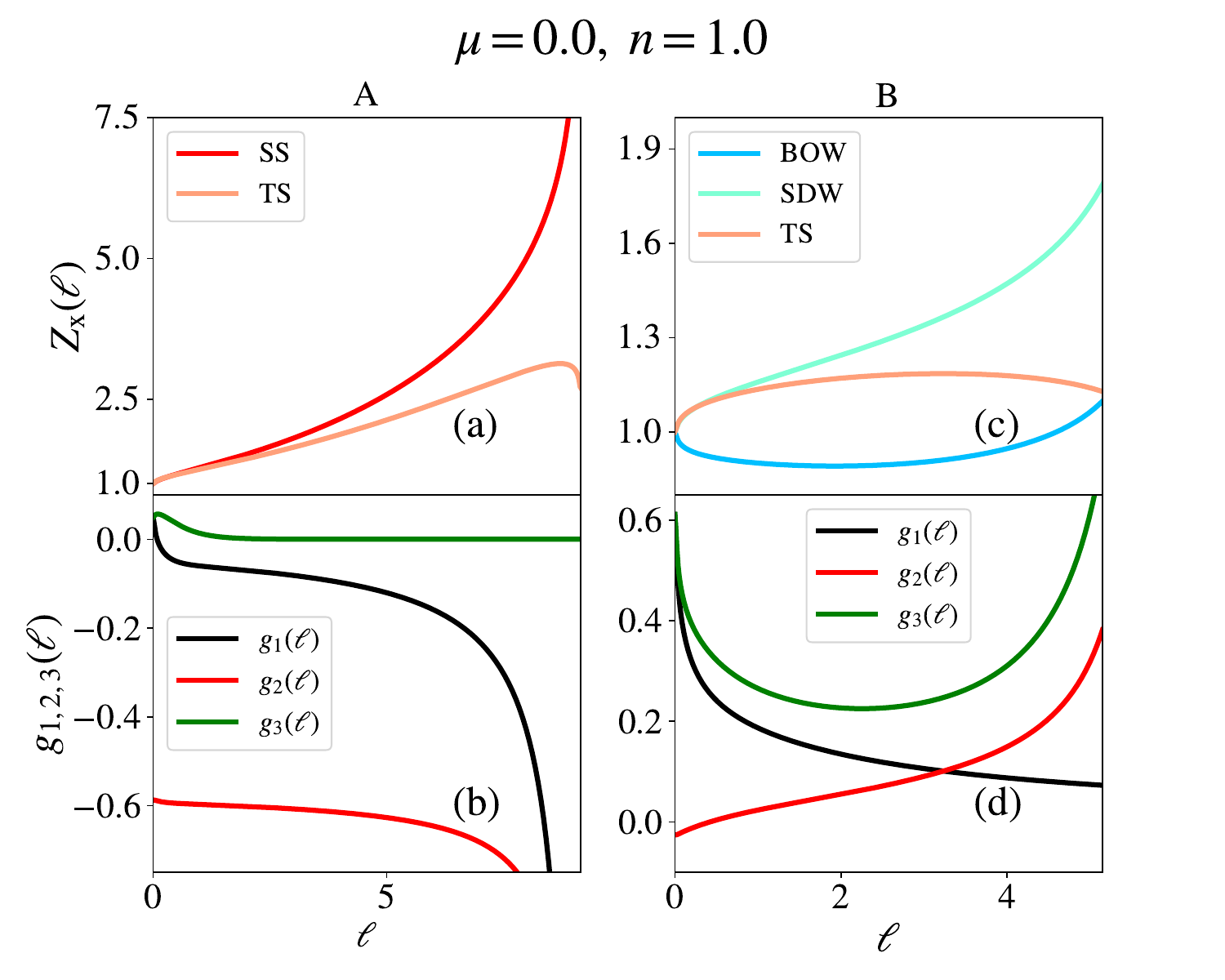}
    \caption{Flow of the three-leg vertices $Z_{\mathrm{x}}$ of the susceptibilities [(a),(c)] and coupling constants [(b),(d)] for points A and B of the phase diagram of  Fig.~\ref{HalfFillingDiagA} at half-filling. A:(-1.69, -1.0), B:(1,84, -1.0).}
    \label{FlowB}
\end{figure}

We now turn to the attractive sector surrounding the $U=2V$ line, namely    the region $U<0$ in the phase diagram of Fig.~\ref{HalfFillingDiagA}. In the $g$-ology formulation of the EHM, the TL conditions $g_1=g_3=0$  at  $U=2V$ will be also unstable    due to  the presence of irrelevant terms that couple  spin and charge degrees of freedom at the beginning of the flow. Thus in spite of $g_3$ remaining irrelevant,  $g_1(\ell)$ becomes negative for $\ell\geq \ell^*$, as shown in Fig.~\ref{FlowB}-(b);  
$\ell^*$ being large, this leads to a small spin gap $\varDelta_{\sigma}$ [Eq.~(\ref{g1star})]. 
As displayed in  Fig.~\ref{HalfFillingDiagA}, this tips the balance in favor of SS as the most stable phase, impinging  on  the region of  TS stability found in the continuum $g$-ology theory (Fig.~\ref{Diag_g_ology}).   The resulting growth of the SS region against the gapless TS one  leads to a convex SS-TS boundary in the phase diagram  that is consistent with previous weak-coupling RG calculations \cite{Menard11} and   exact diagonalization  results of Nakamura \cite{Nakamura00}.  

A  related bending of  phase boundary is also found  for the   $U=-2V$ line separating the Mott SDW and gapless TS phases in the TL model. The SDW state is then favored against TS   at $U>0$ and $V<0$. This is illustrated in Fig.~\ref{FlowB}-(c),(d) for the  point B of Fig.~\ref{HalfFillingDiagA} where $g_2$ changes sign at the beginning of the flow, so that  umklapp  scattering, known to be irrelevant on the $U=-2V$ line in the continuum $g$-ology theory, becomes marginally relevant with a small but finite charge gap in the presence of irrelevant terms, which enlarges  the stability region of the Mott SDW state. 

Regarding the rest of the phase diagram of Fig.~\ref{HalfFillingDiagA}, only  quantitative changes in the flow of coupling constants result from the presence of irrelevant terms due to lattice effects. These results confirm  those of Ref.~\cite{Menard11} obtained by a different RG approach. We close the description of the phase diagram by pointing out the existence of a singularity in the uniform charge compressibility $\chi_\rho$. It signals an instability of the electron system against phase separation which makes an  incursion in the zone of attractive $V$ in  the phase diagram. This incursion is well established by numerical simulations \cite{Nakamura00,Lin95}. Note that for simplicity we didn't include in Fig.~\ref{HalfFillingDiagA} and the following diagrams  at different fillings the continuum prediction for  phase separation.

We  conclude that even if the lattice EHM model at half-filling  is invariably described at sufficiently low  energy by   an effective continuum $g$-ology model, it is difficult to determine the initial conditions of this effective model without a careful analysis of the physics at high energy. Taking directly the continuum limit from the bare Hamiltonian may lead to wrong conclusions as to the nature of the ground state and in turn the structure of the phase diagram. These effects carry over away from half-filling for the EHM model, as we shall discuss next.

\subsection{Away from half-filling}

  As far as the part played by the spectrum is concerned, we first note that away from half-filling, when $\mu\ne 0$, the integration of degrees of freedom is no longer symmetric with respect to the Fermi level,  except in the low-energy domain   where $\varLambda \ll \varLambda_0$ and the spectrum  can be considered  essentially linear, as generically depicted in Fig.~\ref{TightBinding}. 
  
As a consequence, the RG flow can be divided into three regimes. In the first regime, the asymmetry  between electrons and holes plays an important role. Typically, for $ \mu > 0 $, we can have  $ \mathcal{N}(\xi > \La) = 0 $,  that is, no fermion states are available, whereas  $ \mathcal{N}(\xi < - \La) \neq 0 $ (Fig.~\ref{TightBinding}). The bubbles ${\cal L}^{\mathrm{ph,pp}}$  will be affected accordingly. Thus, there will be no  $2\kf$ particle-hole excitations  and ${\cal L}^{\mathrm{ph}}$ will vanish in this regime (See Fig.~\ref{Bubbles}).  This contrasts with Cooper pair excitations,  contributing to   ${\cal L}^{\mathrm{pp}}$, which are present for $(-k,k)$ pairs of momentum  where $\mathcal{N}(\xi_k)\ne 0$. It follows that ${\cal L}^{\mathrm{pp}}$  will be only halved in amplitude, the remaining part being still logarithmic. As we will see, this is responsible for a sizeable screening of interactions at the beginning of the flow, whose impact  alters  the structure of the phase diagram obtained in the continuum limit. This   is reminiscent of the screening of   Coulomb  interactions  by  pairing fluctuations in the theory of conventional superconductivity \cite{Morel62}.  
The second regime corresponds to the $\La$ range where we have $ \mathcal{N}(\xi) \approx \mathcal{N}(-\xi)$, but where the spectrum is still   poorly approximated by a linear function. In this regime, the logarithmic singularity of  the p-h channel is only partly restored while     the one in the p-p channel is complete (See Fig.~\ref{Bubbles}); this imbalance between the two scattering channels favors the screening effects of the Coulomb term.   

 Finally, the last regime corresponds to the continuum limit  at small $\La$, for which we can write $ \mathcal{N}(\xi) \approx \mathcal{N}(-\xi) \approx 1/\pii \vf $. This corresponds to the density of states used in the $g$-ology model for each fermion branch and both spin orientations.

Besides these loop effects associated to the density of states, the chemical potential has also an impact on the Peierls loop ${\cal L}_{\mathrm{P}'} $ in which the reciprocal lattice vector is involved in momentum conservation  making the nesting relation  not perfect anymore \cite{Seidel82,Firsov85}. As a consequence, 
${\cal L}_\mathrm{P}$ survives but not ${\cal L}_{\mathrm{P}'}$, 
so that the equations for normal $g_1$ and $g_2$ processes become independent of umklapp processes at $\La(\ell)< \vf \mu$. 

\begin{figure}
    \centering
    \includegraphics[scale=0.55]{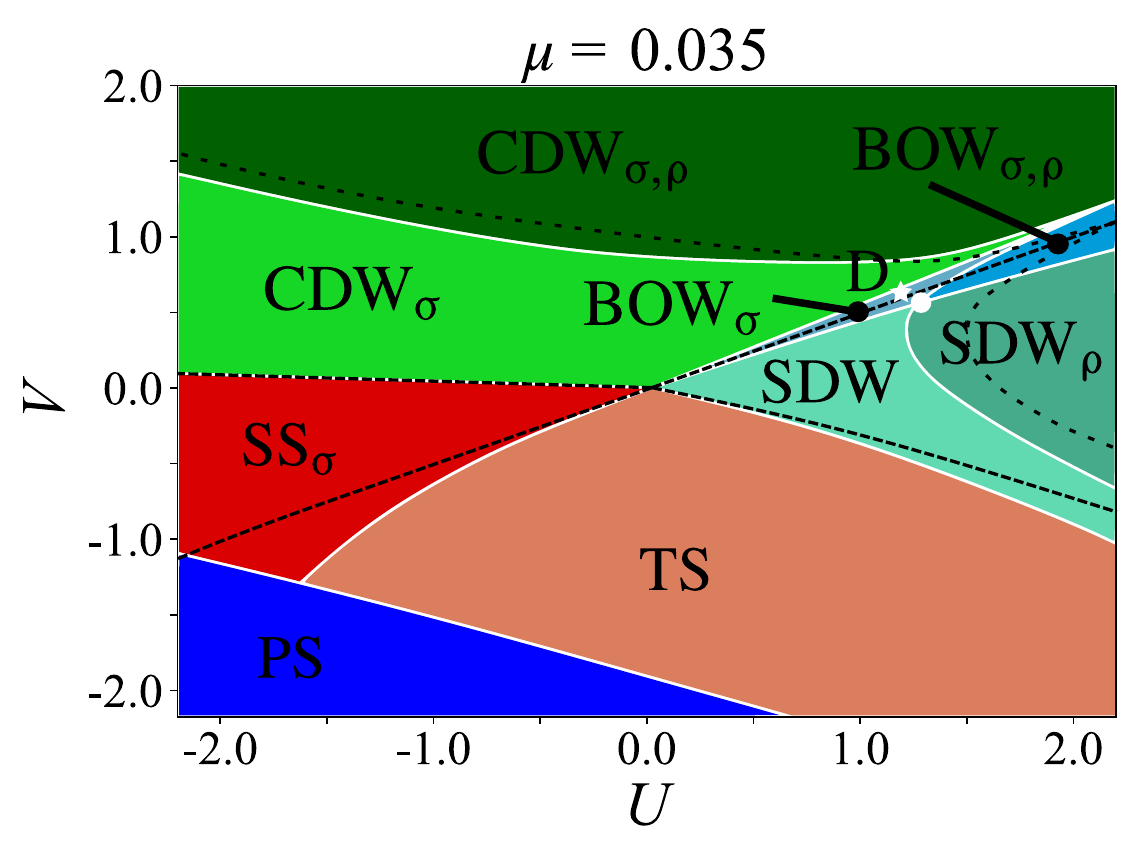}
    \caption{ Same as Fig.~\ref{HalfFillingDiagA} but away from half-filling: $\mu=0.035$.  
    The dashed lines refer to the phase boundaries of the continuum limit in Fig.~\ref{Diag_g_ology}-(b). The open circle corresponds to the threshold value $U_{\mathrm{c}}(\mu)$ for the onset of a gapped BOW state as a function of $\mu$ (Fig.~\ref{Uc}). The point D in the BOW${_\sigma}$  charge-gapless region is discussed in the text and Fig.~\ref{DBOW}.   The long-dashed lines indicate the boundary above which a charge gap is present  in the continuum limit. }
    \label{UVmu_0.035}
\end{figure}

\begin{figure}
    \centering
    \includegraphics[scale=0.6]{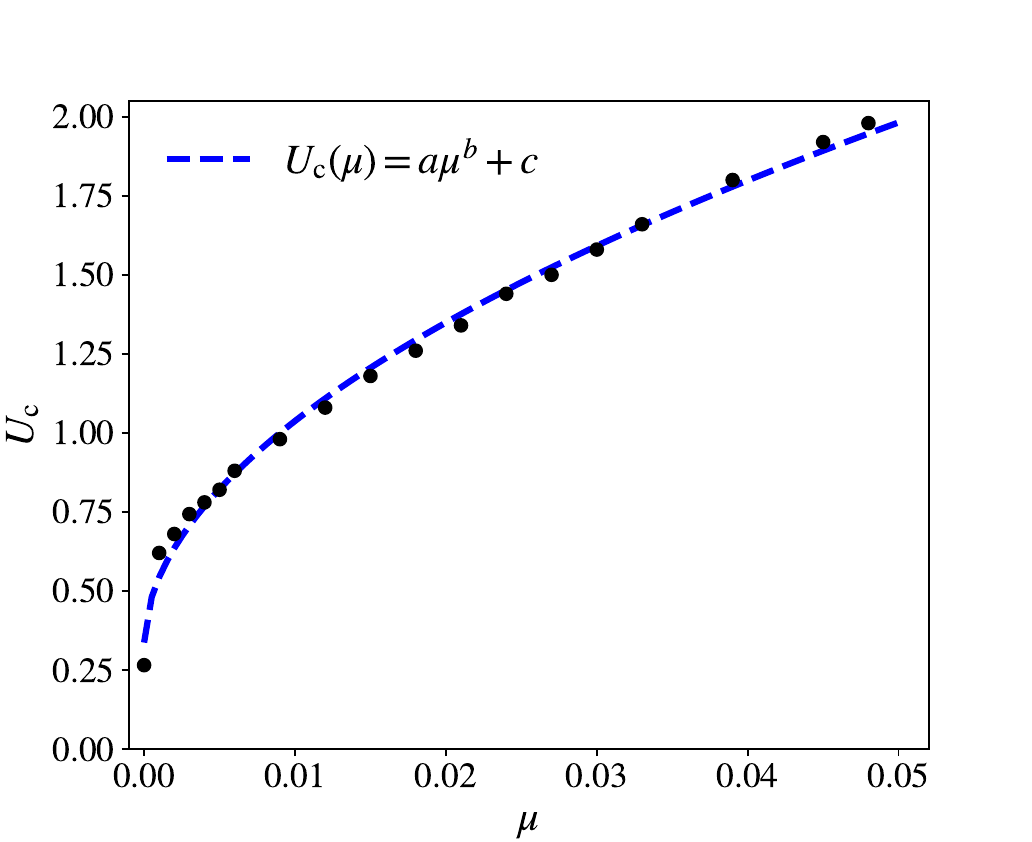}
    \caption{The critical coupling $U_{\mathrm{c}}$ is plotted as a power law $   U_{\mathrm{c}} = a \mu^b + c $ of the chemical potential $\mu$. The gapped   ${\rm BOW}_{\sigma,\rho}$  phase exists for all $ U \geqslant U_{\mathrm{c}} $. Here $ b = 0.53$, $ a = 8.06 $, and the constant  $c = 0.34$ for a temperature of $10^{-7}$ used in the calculations.  }
    \label{Uc}
\end{figure}

\subsubsection{Small doping}

One can now    consider the phase diagram for small departure from half-filling, namely  at $\mu = 0.035$   (Fig.~\ref{UVmu_0.035}), integrating the flow equations~(\ref{gflowlat}), (\ref{zphflowa}-\ref{zphflowb})  and (\ref{zsflow}) with the initial conditions (\ref{gini}), (\ref{gxi0}) and (\ref{gxi20}).   

\begin{figure}[!]
    \centering
    \includegraphics[scale=0.5]{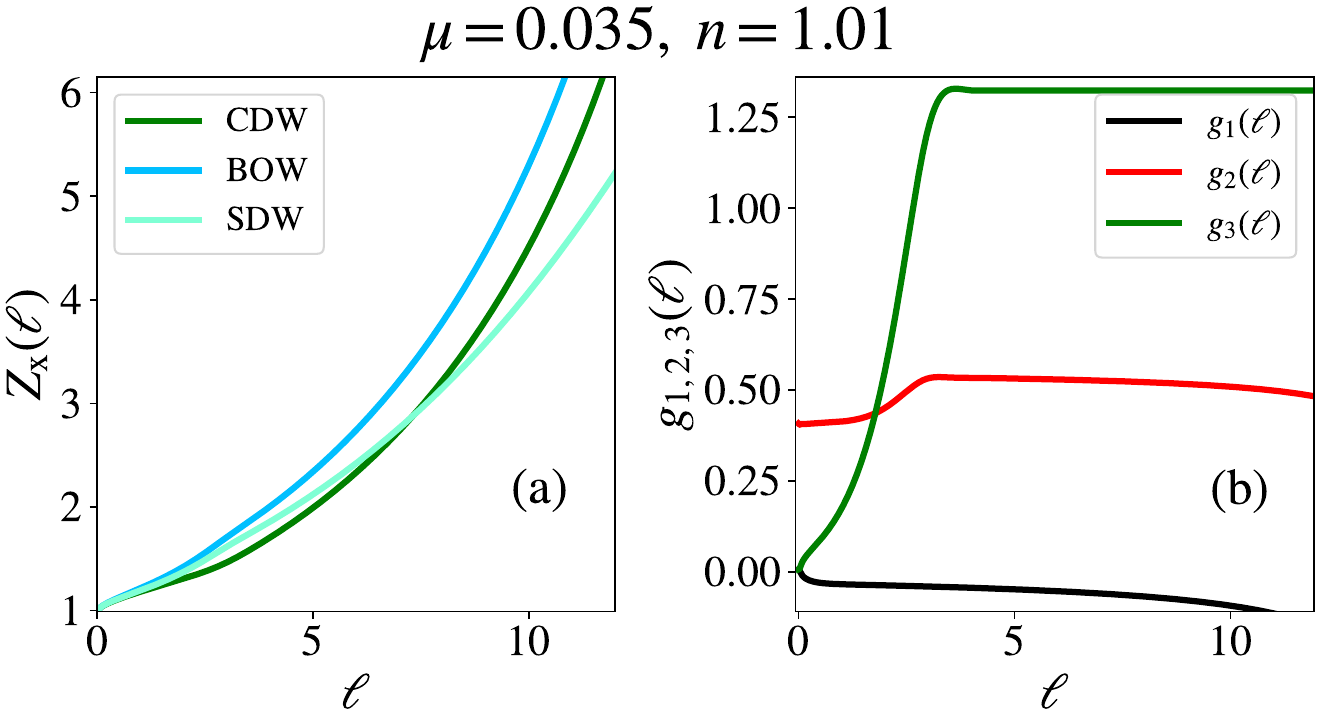}
    \caption{Flow of  the three-leg vertices $Z_{\mathrm{x}}$ of (a) density-wave susceptibilities and (b) coupling constants at point D of the phase diagram of Fig.~\ref{UVmu_0.035} ($\mu=0.035$, $n=1.01$). D:(1.30, 0.63). }
    \label{DBOW}
\end{figure}

In the repulsive sector near the  $U=2V$   line, we see that 
the regions with spin- and charge-gapped BOW and  charge-gapped SDW phases shrink in size, and only exist above some threshold $U_{\mathrm{c}}$ in the interactions. Thus a finite region  unfolds at small coupling   with  CDW, BOW and SDW phases having  no gap in the charge sector (see also footnote \ref{Gapcharge}). The putative gap is  indeed  suppressed by the energy scale $\vf \mu$ that stops the flow of $2g_2-g_1$ and $g_3$ towards  strong coupling when $\La(\ell) < \vf \mu$. The profile of the critical $U_{\mathrm{c}}$ shown in Fig.~\ref{Uc} for the onset of the gapped BOW phase as a function of doping $\mu$, is well described by a power law $ U_{\mathrm{c}}(\mu) \simeq 8.03\mu^{b} + c $, where $b\simeq  0.53$. Here $c\to 0$ when the temperature goes to zero indicating that in the ground state, $U_{\mathrm{c}}\to 0$ as $\mu\to0$. At non-zero $\mu$, a finite region of dominant BOW state with only a spin gap and gapless charge excitations forms in the phase diagram. At point D in Fig.~\ref{UVmu_0.035} for instance,  the corresponding flow  of the couplings displayed in Fig.~\ref{DBOW} shows a growth followed by the leveling off of repulsive umklapp scattering. This is the signature   that $g_3$ becomes irrelevant beyond some finite value of $\ell$. Nevertheless, this trajectory   favors BOW correlations against CDW ones; it also  initiates an incommensurate regime in which $2g_2(\ell)-g_1(\ell)$ evolves toward a constant.  Regarding the attractive backscattering amplitude $g_1$, it will  according to (\ref{g1star})  invariably lead  to a small  spin gap at large $\ell$.

Dominant BOW correlations away from half-filling  but at finite $U$ and $V$ near the line $U=2V$ have been noticed numerically  in  quantum Monte Carlo simulations \cite{Sengupta02},    in qualitative agreement with the present results. If one moves downward in the bottom right quadrant of the phase diagram of Fig.~\ref{UVmu_0.035},  we see that a finite $\mu$ suppresses  the transition for the charge gap at the  boundary between SDW and TS phases which is present at half-filling. The SDW phase then becomes  entirely gapless near the boundary.      
As for the frontier between CDW and SS in top left quadrant of the phase diagram of Fig.~\ref{UVmu_0.035}, it has a bit moved  upward which is consistent with the results of the continuum limit, as already shown in the lower panel of Fig.~\ref{Diag_g_ology}. However, as we will see next  this boundary is noticeably affected at larger $\mu$. 

One finally looks at the singularity line of the uniform charge compressibility $\chi_\rho$. In the attractive $V$ part  of Fig.~\ref{UVmu_0.035}, the  phase separation instability line calculated from (\ref{Runif}) and  the values of $g_\rho^*$ obtained at the end of the flow, undergoes  only a small upward shift with respect to half-filled case.

\subsubsection{Intermediate doping}

\begin{figure}
    \centering
    \includegraphics[scale=0.55]{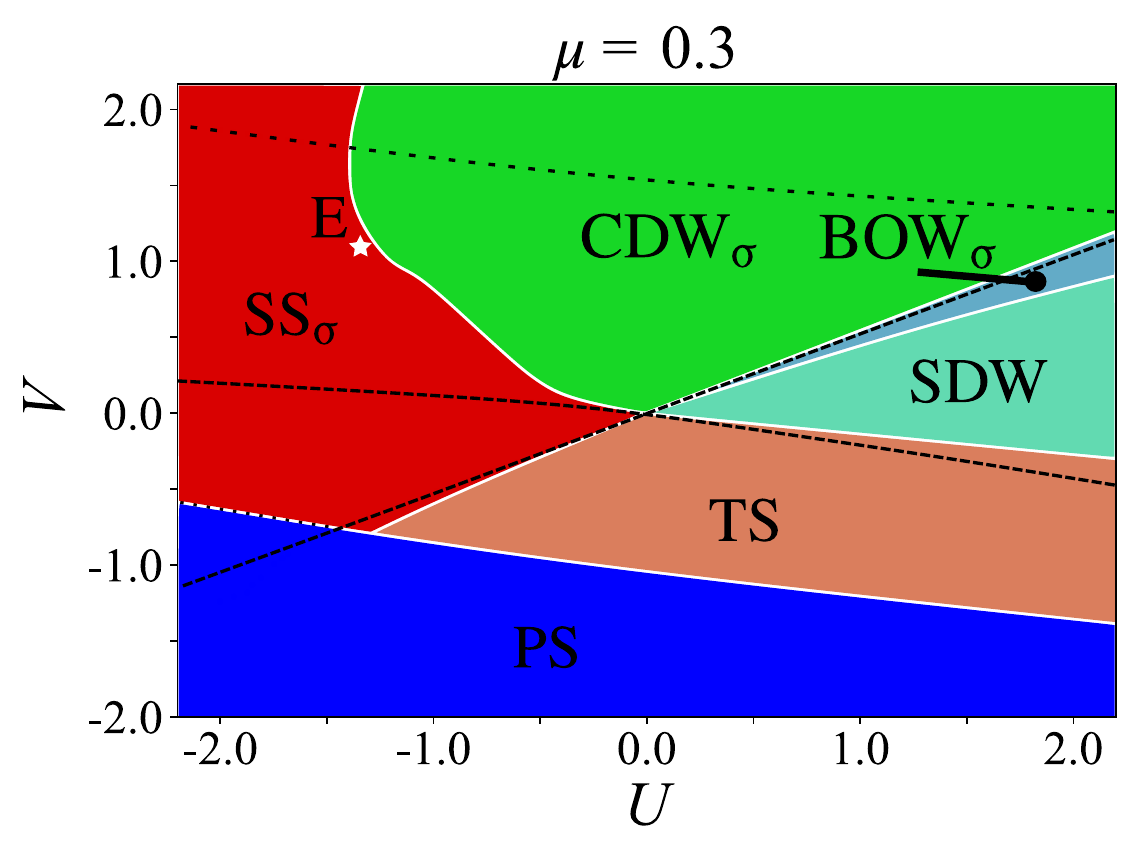}
    \caption{Same as Fig.~\ref{UVmu_0.035} but for  $\mu=0.3$ ($n=1.1$). The point E is discussed in detail in the text. }
    \label{mu03}
\end{figure}

The  phase diagram at intermediate doping $\mu=0.3$ is displayed in Fig.~\ref{mu03}. Due to the  weak effect of umklapp processes at this filling, there is no region of the phase diagram characterized by  a charge gap. However, the influence of $g_3$ at the beginning of the flow  is still finite which, together with the change of $g_1$ to negative values due to irrelevant coupling terms,   still defines near the $U=2V$ line a region of dominant BOW phase at the incommensurate wave-vector $2\kf$. The characteristics of the flow of coupling constants in this BOW region, albeit much further reduced in  their amplitudes, are similar to those shown in Fig.~\ref{DBOW}.      
\begin{figure}[!]
    \centering
    \includegraphics[scale=0.5]{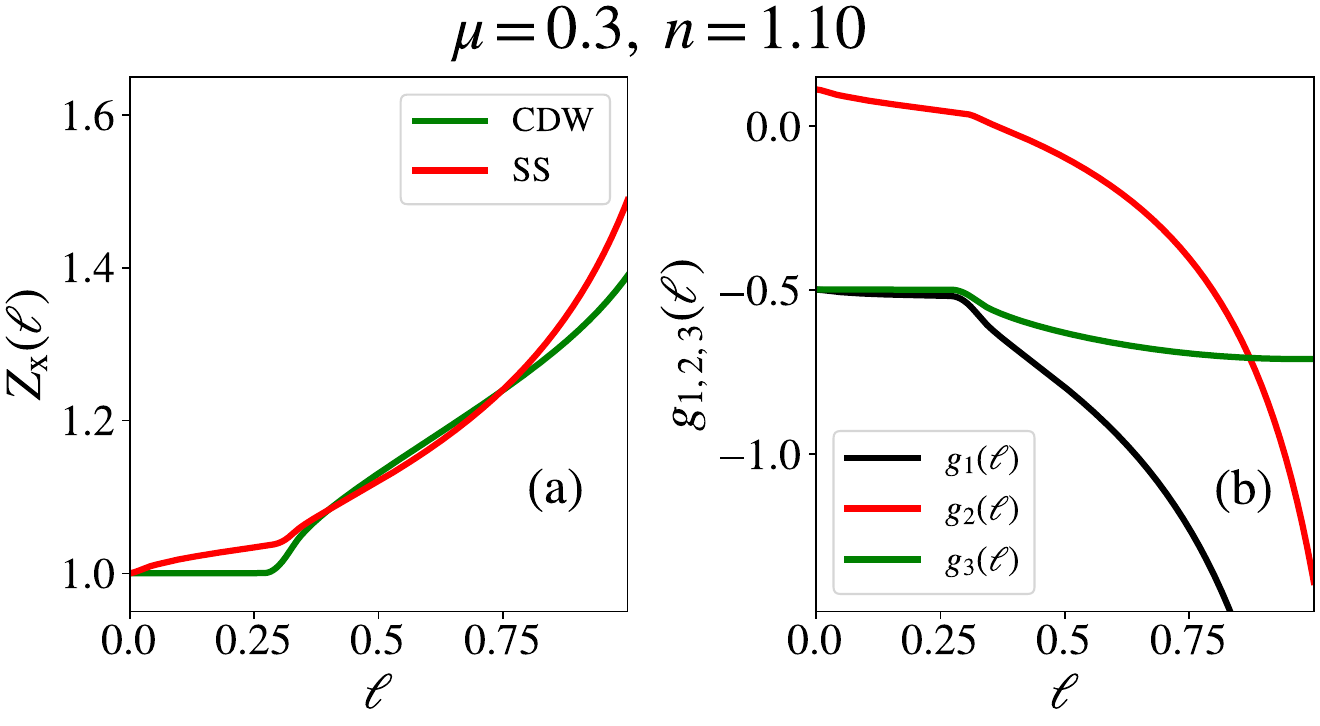}
    \caption{Flow of  (a) the three-leg vertices $Z_{\mathrm{x}}$ of CDW and SS  susceptibilies and (b) the  coupling constants at point E of the diagram  in  Fig.~\ref{mu03} ($\mu = 0.3$,  $n=1.10$). E:(-1.24, 0.97). }
    \label{flowE}
\end{figure}

 In the phase diagram of Fig.~\ref{mu03}, the SDW-TS boundary turns out to be relatively close to the prediction of the model in the continuum limit.  Here only the  weak impact of umklapp and irrelevant couplings,  which  preserves the sign of $g_2$, restores the stability of TS compared to the situation at very small $\mu$ (e.g., point B of Figs.~\ref{HalfFillingDiagA} and \ref{FlowB}-(c),(d)).  

In the top left quadrant of Fig.~\ref{mu03} the deviations with respect to the prediction of the continuum model are particularly significative. One observes an expansion of  the SS phase which goes well beyond  its stability region found in the continuum limit; this occurs   against CDW, which becomes secondary in importance. The origin of this expansion resides in the sizable asymmetry of the spectrum with respect to the Fermi level. At  the beginning of the flow, that is  at large $\La $,  all $2\kf$ particle-hole pair fluctuations coming from   closed loops, 
vertex and ladder diagrams in Fig.~\ref{dgDiag} are strongly suppressed, a consequence of the lack of available density of states for either electrons or holes for this p-h pairing when asymmetry is pronounced, as illustrated in Fig.~\ref{TightBinding}. This regime is followed by a second one at relatively large $\La$ where these fluctuations are only partially restored. Thus there is a sizeable $\La$  interval where p-p ladder diagrams for  pairing fluctuations (first row of Fig.~\ref{dgDiag}) dominate  (see e.g. Fig.~\ref{Bubbles} at finite $\mu$),  and govern the flow of $g_1$ and $g_2$. At point $E$ in Fig.~\ref{mu03} for instance, the coupling $g_2$, though  initially repulsive, is   screened  by pairing fluctuations, to the point where it changes sign and becomes attractive. This is shown in  Fig.~\ref{flowE}-(b). As a result, the SS phase is favored against CDW (Fig.~\ref{flowE}-(a)). This effect is reminiscent of the screening of the Coulomb interaction by  pairing  fluctuations which favors phonon-induced singlet  superconductivity in  isotropic metals \cite{Morel62}.  
The strong reduction of  the $2\kf$ particle-hole pair contribution at the beginning of the flow  is also responsible for  making umklapp  processes  irrelevant in the whole CDW region of the upper half of the phase diagram. This is  why no  charge gap is found, in contrast to the continuum-limit prediction (region above the spaced dashed line in  Fig.~\ref{mu03}). 

Finally we observe on Fig.~\ref{mu03} that the instability line for phase separation undergoes a sizable upward shift to weak coupling values. This is  due  to the smaller  renormalization of $g_\rho^*$ coming from weaker  umklapp scattering and, to a lesser extent, the decrease (increase) in Fermi velocity (density of states). Note that it has not been possible to extract with precision from the flow equations the  $g_\rho^*$ value deep  in the gapped spin region  of the lower left panel of the phase diagram (dashed white line of Fig.~\ref{mu03}). In this region the energy $\varDelta_\sigma \gg T$ at which the flow stops turns out to be  far away from the  thermal energy distance from the Fermi surface where $g_\rho^*$ is defined. In Fig. \ref{mu03} and subsequent phase diagrams, the dashed line corresponds to an extrapolation of the line computed where the spin gap vanishes or is sufficiently close to the thermal scale.

\subsubsection{Large doping}

\begin{figure}
    \centering
    \includegraphics[scale=0.55]{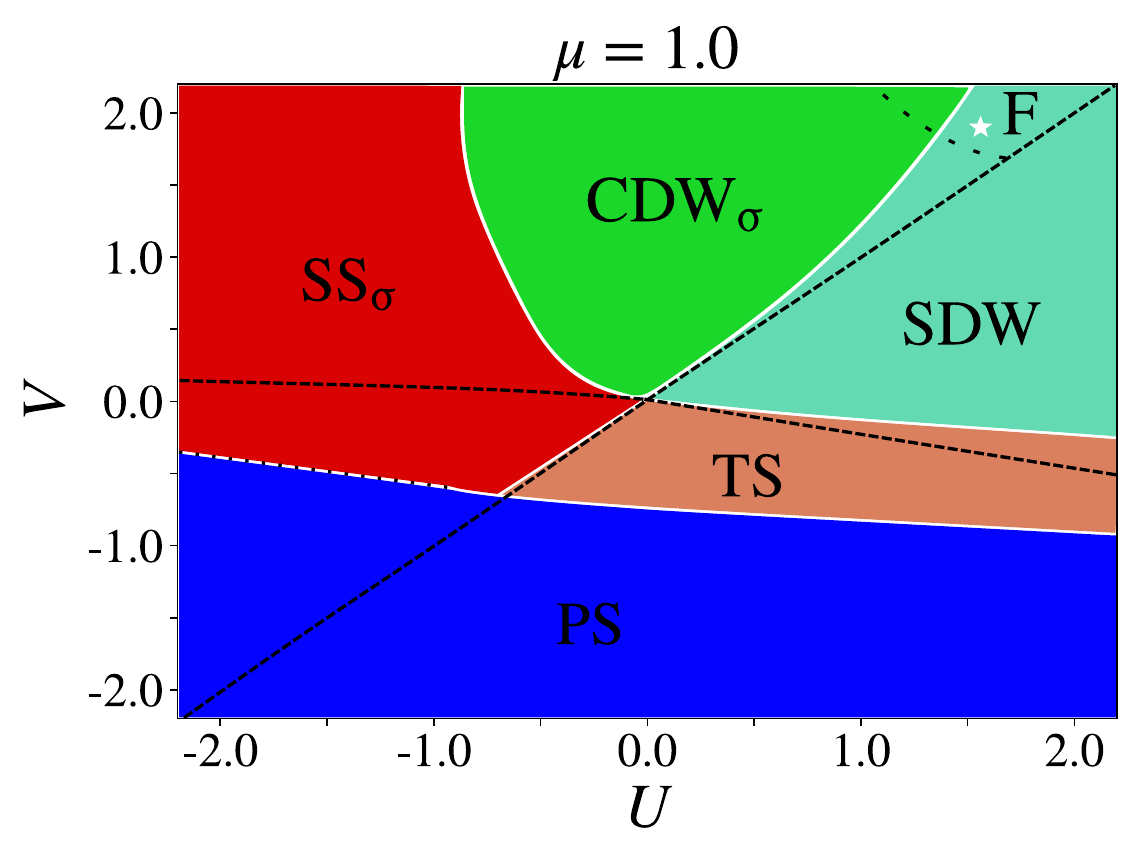}
    \caption{Same as Fig.~\ref{mu03} but for $\mu=1.03$ ($n=1.1$). The point F is  discussed in the text and Fig.~\ref{flowF}.}
    \label{mu1.03}
\end{figure}
One now considers the phase diagram at the higher doping level, ${\mu=1.0}$, shown in Fig~\ref{mu1.03}. The whole diagram indicates  that $g_3$ has virtually no effect in this range of doping reflecting an incommensurate situation for the electron system. This coupling can then be safely ignored in the analysis. Only a spin gap can occur. In the continuum model, we have seen that it  is  governed by the flow of $g_1(\ell)$ [Eq.~(\ref{gologymu})] and the initial condition $ g_1 \simeq U-V<0$ for attractive backward scattering, that is above the dashed line $U\simeq V $ in Fig.~\ref{mu1.03}. According to the figure,  the continuum-model result is however significantly altered by lattice effects and important deviations are present. A significant region develops with gapless spin excitations  although $g_1$ is initially attractive. At point F for instance, Fig.~\ref{flowF} shows that $g_1$ indeed starts in the attractive domain but rapidly evolves towards repulsive sector to become a  marginally irrelevant variable at sufficiently large $\ell$.

This remarkable effect has  its origin  in  the pronounced asymmetry of the spectrum which, as we have seen, suppresses most, if not all, contributions coming from $2\kf$ particle-hole loops at large $\La$;   small momentum pairing fluctuations coming from contributions of ladder Cooper diagrams  to $g_1$ in  Fig.~\ref{dgDiag} largely dominate. Since for these terms   the product $g_1g_2$ in lowest order is initially negative, this makes these diagrams globally positive and pushes the flow of $g_1$ towards positive values. This can be seen as the counterpart effect of the screening discussed above for the enhancement of singlet superconductivity by   pairing fluctuations. This counter-screening of $g_1$ enlarges the  region of gapless spin degrees of freedom in comparison with  the continuum $g$-ology prediction. This in turn  expands the SDW phase at the expense of the spin-gapped CDW phase whose correlations, though still singular, become secondary in importance, as shown in Fig.~\ref{flowF}-(a).  Exact diagonalization studies of the EHM carried out at  $n=2/3\  (\mu=-1.0) $ on the electron-doped side and which corresponds to $n=4/3\  (\mu =1.0)$ in the  hole-doped case of Figs.~(\ref{mu1.03}-\ref{flowF}),  have clearly identified such corrections to the spin gap line of the continuum $g$-ology approach  \cite{Sano94}. However, in this enlarged  region with no  spin gap, the superconducting TS and SS susceptibilities are not enhanced with respect to the free electron limit.

\begin{figure}[!]
    \centering
    \includegraphics[scale=0.5]{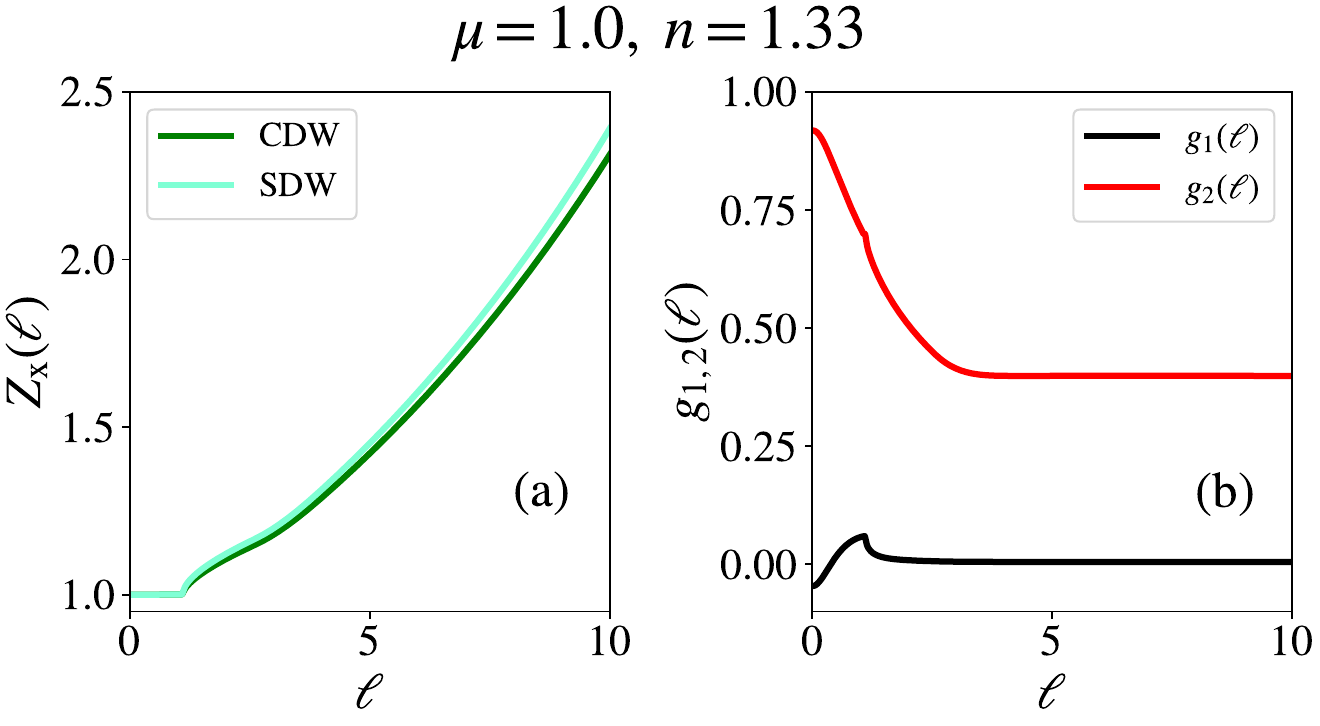}
    \caption{Flow of the three-leg vertices $Z_{\mathrm{x}}$ of the  susceptibilities (a) and couplings (b) at point F of Fig.~\ref{mu1.03} at $\mu =1.0~(n=1.33)$ . F:(1.50, 1.75).}
    \label{flowF}
\end{figure}

\begin{figure}
    \centering
    \includegraphics[scale=0.55]{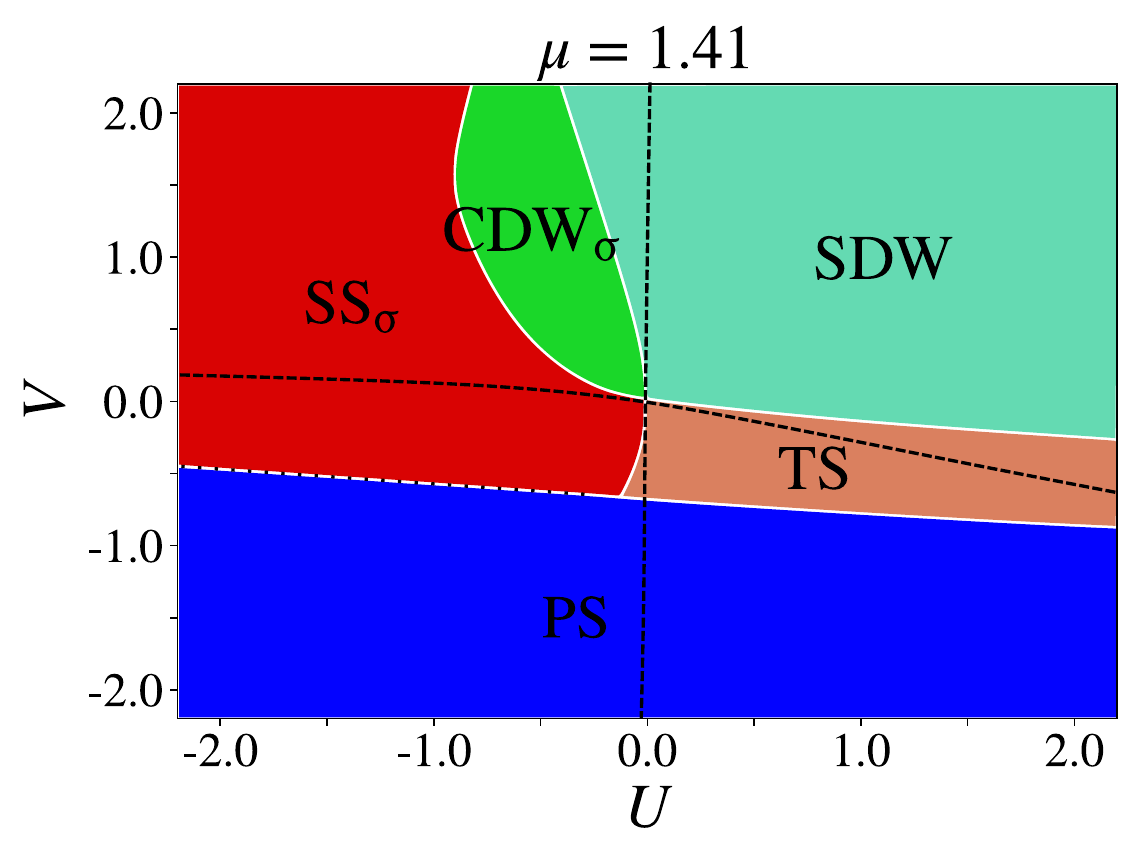}
    \caption{Phase diagram of the EHM at $3/4$-filling $\mu =\sqrt{2}\  (n=1.5)$.}
     \label{quarterfilling}
\end{figure}

In the attractive $V$ region the phase separation line continues to be slightly  shifted upward to weaker coupling due to more favorable initial values $g_\rho$ and lower $\vf$. As in the previous case with $\mu = 0.3 $, the values of $g_\rho^*$ cannot be accurately extracted from the flow in the spin gapped region (dashed white line of Fig.~\ref{quarterfilling}), where the flow stops at an energy scale  far from $T$ and thus away from the conditions of the continuum limit.

If we  turn our attention to the top left quadrant of the phase diagram, we see that compared to the results shown in Fig.~\ref{mu03} the  stability region of  SS phase is further broadened  against the CDW one of smaller amplitude. The SS region reaches about twice the area predicted by the continuum model. The screening of $g_2$ by  pairing fluctuations from positive to negative values results from the  non-linear spectrum  in the first two regimes  
of the flow. This  follows  the  pattern already displayed in Fig.~\ref{flowE}, which  is here magnified  due to the more pronounced  asymmetry of the spectrum. This trend is confirmed  when $\mu$ is further increased. 

This is illustrated by the calculations performed at the higher commensurate  doping  $\mu=\sqrt{2}$ ($3/4$-filling, $n=1.5$). These yield the phase diagram   shown in Fig.~\ref{quarterfilling}  which is roughly  similar  to Fig.~\ref{mu1.03}, except for the boundaries delimiting   the CDW phase. By the same mechanism of screening the CDW region   monotonously  shrinks in size showing sign of closing at stronger coupling, this to the benefit of the SS or SDW phase. On the SDW side, this is  concomitant with the expansion of gapless region for spin degrees of freedom due to counter screening at the beginning of the flow.  Results of exact diagonalisations at quarter-filling are congruent with these corrections \cite{Penc94,Sano94}.\footnote{The present fRG calculations do not take into account the influence of the   $8\kf$ umklapp scattering which involves the transfer of four particles from one Fermi point to the other and that is present at $3/4$ (or $1/4)$ filling \cite{Giamarchi04}. This higher-order umklapp scattering is $\mathcal{O}(\xi^2)$ in power counting and  is thus irrelevant. It is known to only affect qualitatively the phase diagram  beyond some critical $V>0$ value where  a $4\kf$ charge ordered state  is found \cite{Mila93,Sano07}. This regime is well outside the weak coupling sector considered in this work. However, one cannot exclude  that it  may affect the flow of marginal couplings.} Regarding the instability line of phase separation at attractive $V$, only a small upward shift in its  position  is found with respect with the previous  $\mu =1.0$ or $n=1.3$ situation owing to the slight increase  in  the initial value of $g_\rho$ and  of the density of states.   

\begin{figure}
    \centering
    \includegraphics[scale=0.55]{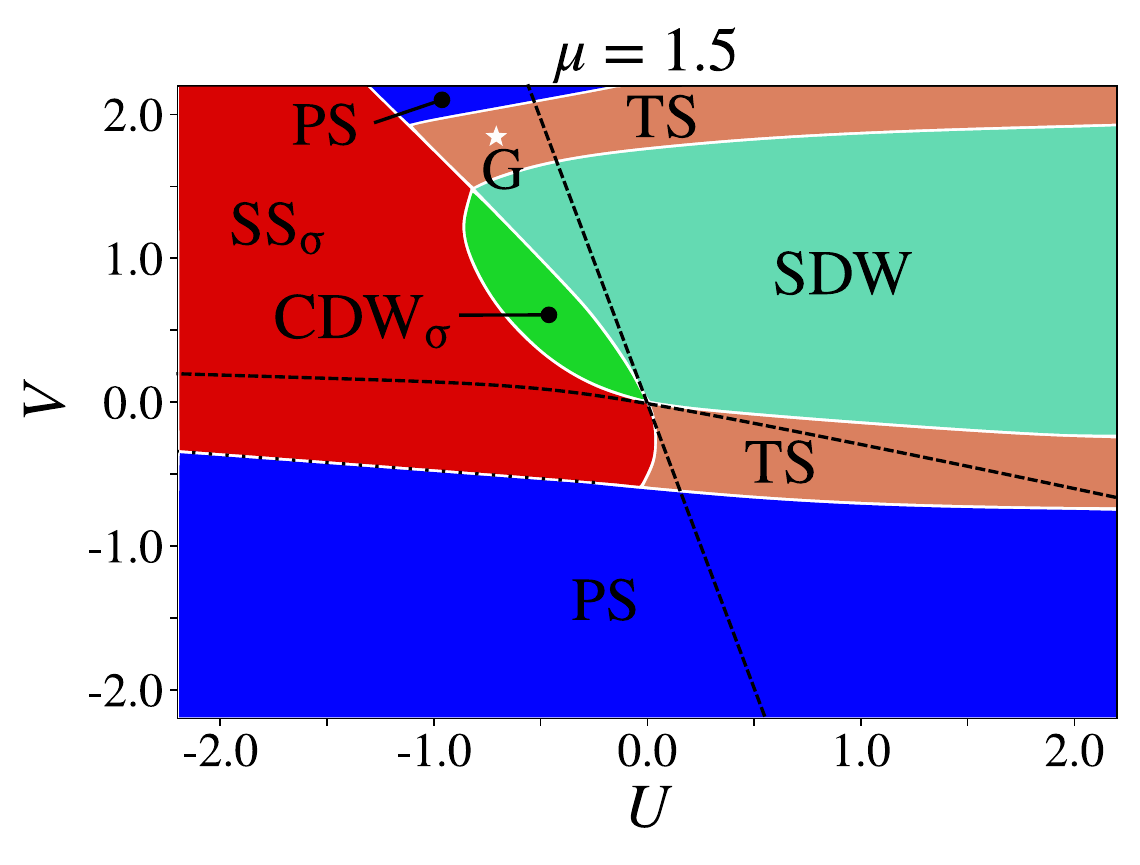}
    \caption{Phase diagram of the EHM at large $\mu=1.5$ $(n=1.54)$.}
    \label{mu1.5}
\end{figure}

When the doping is  further increased beyond the 3/4-filling, qualitative changes in the phase diagram become manifest. As shown in Fig.~\ref{mu1.5} for $\mu=1.5 \ (n=1.54 )$, the region of gapless spin degrees of freedom continues to be enlarged with respect to the one of the continuum $g$-ology results, but the most striking result resides in the closing of the CDW zone at $V>0$ which gives way to  the  emergence of a TS phase with gapless excitations in the spin sector. This result contrasts with what is found in the $g$-ology framework where the TS phase is confined to the lower right part of the phase diagram.  As shown in Fig.~\ref{flowF}-(b), the  counter screening  of $g_1$ ($g_2$) to positive (lower) values at the beginning of the flow is responsible for the occurrence of gapless superconductivity in this part of the phase diagram. As displayed in Fig.~\ref{flowF}-(a), only TS correlations  are singularly enhanced in this region while those of the SS type are reduced  compared to the free electron limit. 
 
To our knowledge no numerical simulations have been carried out at this doping which would allow a precise comparison. However,  results of exact diagonalizations  at quarter filling ($\mu = -1.41$) have revealed the existence of a peculiar and unexpected superconducting phase in the gapless sector for $V \gtrsim 4$ and $U$ nearly centered around zero \cite{Penc94,Sano94}. This would be located  above the weak-coupling CDW  region of Fig.~\ref{quarterfilling}. The present results  strongly suggest that superconductivity found by exact diagonalizations  at 1/4-filling corresponds to the TS phase  that emerges in  Fig.~\ref{mu1.5} at a smaller $V>0$.  Finally, the phase separation line in the attractive $V$ domain continues its slow upward shift, another instability  line of this type begins to appear, but this time in the  TS region described above at repulsive $V$. It is worth mentioning  that at larger positive $V$ numerical simulations achieved  at 1/4-and 2/3-fillings also find such a phase within the previously described TS region \cite{Penc94,Sano94}.   

\begin{figure}[!]
    \centering
    \includegraphics[scale=0.5]{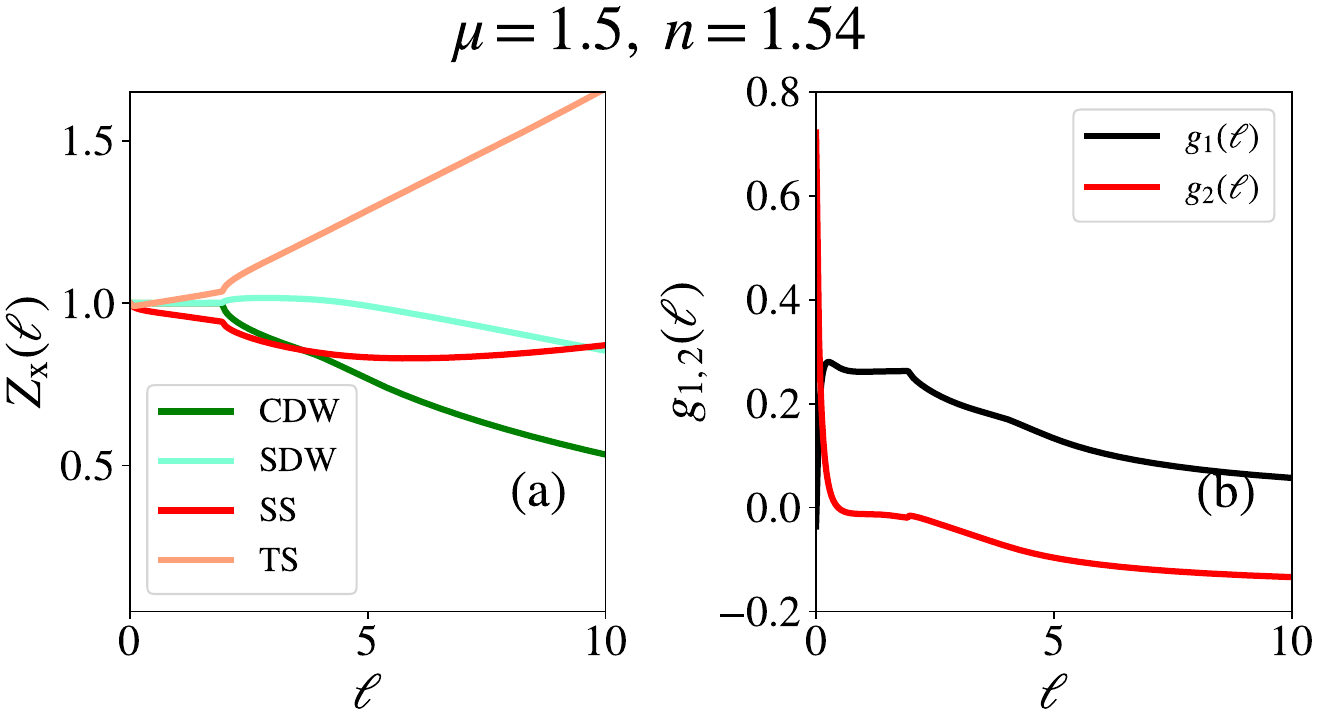}
    \caption{ Flow of the three-leg vertices $Z_{\mathrm{x}}$ of the  susceptibilities (a) and couplings (b) at point G of Fig.~\ref{mu1.5} at $\mu =1.5~(n=1.54)$. G:(-0.60, 1.80).}
    \label{flowF}
\end{figure}

From the above results one can conclude   that the asymmetry between occupied and unoccupied electron   states in  an incommensurately filled spectrum  can introduce pairing fluctuations which act as an efficient mechanism to modify  the repulsive part  of long-range  Coulomb  interactions at low energy  and  thus promote superconductivity of different nature.

\section{Conclusions and perspectives}
\label{conclusion}

In this work we have developed a weak-coupling functional  RG approach to  1D   lattice  models of interacting fermions in one dimension.  In the framework of the EHM, we have shown how lattice effects modify in a systematic way the initial conditions defining the effective  continuum field theory which invariably  emerges  at sufficiently  low energy. For repulsive couplings at half-filling, for instance, the impact of  irrelevant interactions   on marginal couplings, which couple spin and charge degrees of freedom,  turn out to be a key factor in the emergence of the  gapped  BOW state that overlaps the $U=2V>0$ gapless TL line of the   continuum theory.  We have also checked that qualitative changes in the nature of ground states are also manifest in the attractive  sector of  the EHM phase diagram at half-filling.  These changes are due to irrelevant terms affecting the flow of marginal couplings at high energy and introducing  noticeable shifts in the  transition lines of the continuum theory, altering the stability region of the gapless TS state in favour of  SS or SDW gapped states. These alterations of the continuum EHM phase diagram at weak coupling  are consistent  with previous numerical  studies \cite{Nakamura99,Nakamura00}; they also confirm  the results obtained from numerical fRG in the repulsive coupling  sector \cite{Tam06}, and more generally from a   Wilsonian RG approach to the non linearity of the spectrum and momentum-dependent interaction  of the EHM \cite{Menard11}. 

We have also carried out our fRG procedure away  from half-filling.  In this case, the particle-hole symmetry in the tight-binding spectrum  is lost and the integration of degrees of freedom becomes asymmetric with respect to the Fermi level. This notably affects   the influence of high-energy fermion states on  the flow  of scattering amplitudes and susceptibilities. An imbalance between the logarithmic screening of the p-p  and $2\kf$ p-h  scattering channels is introduced which couples charge and spin degrees of freedom. In a finite energy interval at the beginning of the flow, the $2\kf$ density-wave part and concomitantly the magnification of umklapp commensurability, are strongly reduced.  This contrasts with   the  p-p scattering channel which is weakly affected and  sees its logarithmic  singularity maintained. As the integration of degrees of freedom approaches the Fermi level, the  imbalance together with irrelevant interactions scale down to zero and the flow progressively evolves toward the one of an effective continuum theory.   However, the   input parameters that govern  the low-energy flow  are not those of the naive continuum limit  and  alter sizable parts of the EHM  phase diagram compared to the continuum $g$-ology predictions away from half-filling.  This is particularly manifest for the CDW  state    whose  extent in the phase diagram as the dominant phase at negative $U$ and repulsive $V$, for instance,  is steadily reduced as a function of  doping to the benefit of singlet   superconductivity which gains in importance.  This  feature is not without bearing comparison with the screening of Coulomb interactions  by high-energy  pairing fluctuations in ordinary metals, which is known to promote the existence of superconductivity from retarded attractive coupling induced by  electron-phonon interactions \cite{Morel62}. At large doping and repulsive $V$, pairing fluctuations are found to promote repulsive back scattering interactions by expanding the region of gapless spin excitations. This occurs with  the emergence of a TS phase compatible with the one found by exact diagonalization studies carried out at stronger coupling  far from half-filling \cite{Sano94,Penc94}.

The approach developed in this paper can be easily transposed to other non-integrable  models of interacting electrons defined on a lattice. This is  the case of models  with generalized non-local interactions \cite{Kivelson87,Campbell88, Campbell90,Japaridze99}, for which numerical calculations are available at half-filling  and  known to deviate from the predictions of the   $g$-ology approach in the field-theory continuum limit \cite{Nakamura00}. Another natural extension of the present work concerns the EHM in the quasi-one-dimensional case,  where a weak but finite interchain hopping is taken into account. This may serve as a weak-coupling quasi-1D EHM to study the sequence of ground states that can unfold in strongly anisotropic correlated systems as  a function of doping.  Some of these applications are currently under investigation.

C.~B  and L.~D thank  the National Science and Engineering Research Council  of Canada (NSERC), the Regroupement Qu\'eb\'ecois des Mat\'eriaux de Pointe (RQMP) and the Institut Quantique of Université de Sherbrooke for financial support.  The authors thank E. Larouche and M. Haguier for their  support  on various numerical aspects of this work.

\appendix

\section{Flows  of coupling constants}

\subsection{Finite-temperature, one-dimensional, single-band systems}
\label{Appendix}

In this first part of the Appendix, we detail the derivation of the flow equations for the scattering amplitudes at the one-loop level for both marginal and  irrelevant couplings.  To do so we first make the correspondence $k\to (\eta,\xi)$   between the momentum and  the energy $\xi$  and its branch $\eta$, so that 
$$
g_{k_1, \, k_2, \,  k'_1}  = g^{\vec \eta}(\vec \xi),
$$
where  $\vec x  = (x_1, x_2, x_{1'})$  for  $ x = \xi, \eta$. 
From  the diagrams of Fig.~\ref{dgDiag}, the flow equations of the coupling constants   at the one-loop level comprise a sum of contributions coming from p-p and p-h scattering channels, which can be put in the form:
\begin{align}
\La \dr_\La g^{\vec \eta}(\vec \xi) & = \sum_{\mathrm{x}} D^{\vec \eta}_{\mathrm{x}}(\vec \xi) \cr
&= D^{\vec \eta}_{\mathrm{pp}}(\vec \xi) + D^{\vec \eta}_{\mathrm{ph1}}(\vec \xi) + D^{\vec \eta}_{\mathrm{ph2}}(\vec \xi) + D^{\vec \eta}_{\mathrm{ph3}}(\vec \xi) , 
\label{FlowG}
\end{align}
where the diagrams 
\begin{equation}
     D^{\vec \eta}_{\mathrm{x}}(\vec \xi) =\, \sum_p \, \mathcal{L}^{\vec \eta}_{\mathrm{x}} (\vec \xi) \, \gamma^{\vec \eta}_{\mathrm{x}1} (\vec \xi) \,  \gamma^{\vec \eta}_{\mathrm{x}2} (\vec \xi)
\end{equation}
are expressed in terms of loops $\mathcal{L}^{\vec \eta}_{\mathrm{x}} (\vec \xi)$ and combinations of coupling constants $\gamma^{\vec \eta}_{\mathrm{x}1} (\vec \xi)$ and $\gamma^{\vec \eta}_{\mathrm{x}2} (\vec \xi)$ for each scattering channel $\mathrm{x}$. They are respectively given  by 
\begin{subequations}
\begin{align}
\label{Dee}
 D^{\vec \eta}_{\mathrm{pp}}(\vec \xi) = & \, \sum_p \, \mathcal{L}^{\mathrm{pp}}_{p, \, -p+k_1+k_2} \,  g_{k_2, \, k_1, \, -p+k_1+k_2} 
 g_{p, \, -p+k_1+k_2, \, k'_1} , \\
\label{Deh1}
D^{\vec \eta}_{\mathrm{ph1}}(\vec \xi) = &\, \sum_p \, \mathcal{L}^{\mathrm{ph}}_{p, \, p-k'_1+k_2}\, g_{k_1, \, p-k'_1+k_2, \, p} 
g_{k_2, \, p, \, p-k'_1+k_2}, \\
\label{Deh2}
D^{\vec \eta}_{\mathrm{ph2}}(\vec \xi) = &- 2 \sum_p \,   \mathcal{L}^{\mathrm{ph}}_{p, \, p+k'_1-k_1}\, g_{k_1, \, p+k'_1 - k_1, \, k'_1} 
g_{k_1, \, p+k'_1 - k_1, \, k'_1} \,   g_{p, \, k_2, \, p+k'_1 - k_1} , \\
\label{Deh3}
D^{\vec \eta}_{\mathrm{ph3}}(\vec \xi) = & \,\sum_p \, \mathcal{L}^{\mathrm{ph}}_{p, \, p+k'_1-k_1}  \, (  g_{k_1, \, p+k'_1 - k_1, \, p} \,   g_{p, \, k_2, \, p+k'_1 - k_1}  
+  g_{k_1, \, p+k'_1 - k_1, \, k'_1} \,   g_{k_2, \, p, \, p+k'_1 - k_1}   ) .
\end{align}
\end{subequations}
As explained in the main text,  $ g^{\vec \eta}(\vec \xi)$ and $   D^{\vec \eta}_{\mathrm{x}}(\vec \xi)$  on each side of (\ref{FlowG}) can be formally expanded in power of $\vec \xi$ to get the flow equations of the set of marginal and irrelevant couplings.

\subsection{Loop expressions}

In order to derive the expressions of the bubble intensities, let us first introduce the free propagator regularized at scale $ \La $:  
\begin{equation}
    G^\La_0(p_n, p) = \dfrac{\theta_\Lambda(p)}{\ii p_n - \xi(p)},
    \label{Reg_propagator}
\end{equation}
where $p_n$ denotes the fermionic Matsubara frequencies and $p$ the momentum. The loop expressions are then obtained from the derivative of the product of the propagators:
\begin{equation}
    \mathcal{L}_{(p_n,p),(q_n,q)} = - \dfrac{\pii \vf T}{L} \La \dr_\La \big( G^\La_0(p_n, p) G^\La_0(q_n, q)  \big).
\end{equation}
and the sum over the Matsubara frequencies is then performed:
\begin{equation}
    \begin{split}
    \mathcal{L}^{\mathrm{pp}}_{p,q} & = \sum_{p_n}    \mathcal{L}_{(p_n, p),(-p_n,-p+q)} , \\
    \mathcal{L}^{\mathrm{ph}}_{p,q} & = \sum_{p_n}  \mathcal{L}_{(p_n, p),(p_n,p+q)}.
\end{split}
\end{equation}
The expressions of the  loop contributions for the diagrams of the $\mathrm{p-p}$ and $\mathrm{p-h}$ scattering channels are thus given by
\begin{equation}
    \begin{split}
\label{Loops}
    \mathcal{L}^{\mathrm{pp}}_{p,q} & = - \dfrac{\pii \vf }{2 L} \La \dr_\La \big( \theta_\La(p) \theta_\La(p+q) \big) \dfrac{\FD\big(\xi(p) \big) - \FD\big(-\xi(p+q) \big)}{\xi(p)+\xi(p+q)}  ,  \\
    \mathcal{L}^{\mathrm{ph}}_{p,q} & = \dfrac{\pii \vf }{2 L} \La \dr_\La \big( \theta_\La(p) \theta_\La(p+q) \big) \dfrac{\FD\big(\xi(p) \big) - \FD\big(\xi(p+q) \big)}{\xi(p)-\xi(p+q)} .
\end{split}
\end{equation}

where $ \FD(\xi) = (1+\ee^{\beta \xi})^{-1}  $ is the Fermi-Dirac distribution  and $ \theta_\La(k)$ is the regulator or cut-off function of the RG procedure. The latter  is introduced explicitly in Sec.~\ref{Regul} below.

\begin{figure}[!]
    \centering
    \includegraphics[scale=1.25]{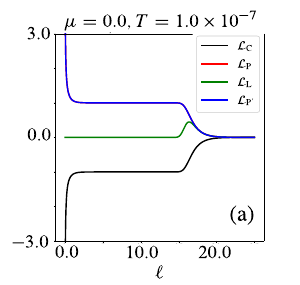}
    
    \includegraphics[scale=1.25]{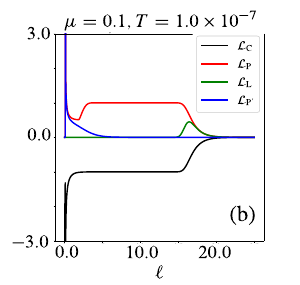}
    
    \includegraphics[scale=1.25]{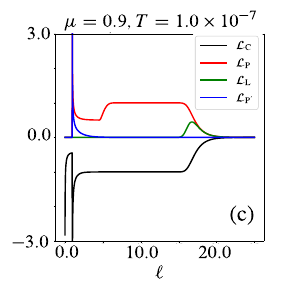}   
    \caption{
    Cooper, Peierls and Landau bubbles  shown for different values of the chemical potential, in the case of a tight-binding spectrum. The first panel is at half-filling and the others at different fillings.  In this figure, $ \ell $ is the RG time, defined by  $ \La = \La_0 \ee^{-\ell} $. The sharp peaks appear when $\La$ hits the band edges and are due to the van Hove singularity in the density of states. }
    \label{Bubbles}
\end{figure}

Let us discuss some limiting cases for these loops at vanishing external momentum. These  enter in the flow equations of  response functions.  We can define the following intensities in each scattering channel. In the p-h channel, we have
\begin{equation}
\begin{split}
\mathcal{L}_{\mathrm{P}} & =  \sum_{p\geqslant 0} \mathcal{L}^{\mathrm{ph}}_{p, \, p- 2\kf} = \sum_{p\geqslant 0}  \mathcal{L}^{\mathrm{ph}}_{-p, \, -p+ 2\kf}, \\
\mathcal{L}_{\mathrm{P}'} & =   \sum_{p\geqslant 0}   \mathcal{L}^{\mathrm{ph}}_{p, \, p+ 2\kf} = \sum_{p\geqslant 0}  \mathcal{L}^{\mathrm{ph}}_{-p, \, -p - 2\kf}, \\
\mathcal{L}_{\mathrm{L}} & =  \sum_{p\geqslant 0}   \mathcal{L}^{\mathrm{ph}}_{p, \, p} =  \sum_{p\geqslant 0}  \mathcal{L}^{\mathrm{ph}}_{-p, \, -p},
\end{split}
\end{equation}
which  correspond respectively to the $2\kf$ p-h or Peierls loops without ($\mathcal{L}_{\mathrm{P}}$) and with ($\mathcal{L}_{\mathrm{P}'}$) umklapp scattering, and to $q=0$ p-h loop.  
As for the p-p  or Cooper loop at zero pair momentum, it is given by
\begin{align}
\mathcal{L}_{\mathrm{C}} & =  \sum_{p\geqslant 0}  \mathcal{L}^{\mathrm{pp}}_{p, \, -p} =  \sum_{p\geqslant 0} \mathcal{L}^{\mathrm{pp}}_{-p, \, p}.
\end{align}
These quantities are plotted in  Fig.~\ref{Bubbles} as a function of the RG time $ \ell $ defined by $ \La = \La_0 \ee^{-\ell} $. We can observe the presence of the van Hove singularity located at the edge of the spectrum. 
At half-filling the amplitudes of the Cooper and  Peierls  bubble intensities  are the same at all $\ell$ but opposite in sign, and lead to maximum interference between certain classes of diagrams in Fig.~\ref{dgDiag}. Away from half-filling,  the Peierls intensity   $\mathcal{L}_{\mathrm{P}'}$, which  involves umklapp scattering,  sees its   intensity suppressed as a function of $\ell$, when typically $\Lambda(\ell)< \vf \mu$. This differs from the normal part $\mathcal{L}_{\mathrm{P}}$  with no umklapp scattering which keeps its full intensity  down to the thermal shell.  We also note in the third panel of Fig.~\ref{Bubbles} that at sizeable doping  all the Peierls intensities are zero at the beginning of the flow. This results from the particle-hole asymmetry of the spectrum which suppresses  electron or hole  states required for $2\kf$ p-h pairing. By contrast the asymmetry of the spectrum suppresses only half of the states for p-p pairing states so that the Cooper intensity is only halved and remains finite at the beginning of the flow.      

\subsection{Renormalization of the Fermi velocity}

\label{Fermi_velocity}

Let us compute the Hartree-Fock contributions to the renormalization of the one-body term. They can be put in the diagrammatic form
\begin{align}
\dr_\La \varSigma_{k} & =  \raisebox{-1pt}{\includegraphics[scale=0.32]{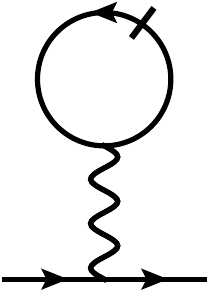}} +{\includegraphics[scale=0.32]{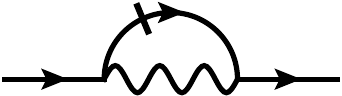}}  \raisebox{-0.4pt}.
 \label{1Pself}
\end{align} 
One can expand $\varSigma_k\simeq \varSigma_0 + \varSigma_1\xi(k) $ to first order in the  energy $\xi$. The momentum-independent term  $\varSigma_0$ renormalizes the chemical potential.
However, it can be rescaled back to its initial value (\ref{mu}) at each step of the flow for a given band-filling, so it 
is the bare value of $\mu$ that is used in the flow equations. The momentum-dependent term linear in $\xi$, $\varSigma_{1,\La}$, leads to the flow of the hopping term $t_\Lambda$ or correspondingly of the renormalization of the Fermi velocity $v_{\mathrm{F}\La}=\vf(1+ \varSigma_{1,\La})$.

From the evaluation of the second Fock term of (\ref{1Pself}), one has using a sharp cutoff
\begin{align}
    \partial_\La v_{\mathrm{F}\La} =&{} \,\frac{\vf^2}{8}   \int_{-\La_0}^{+\La_0}  \dd \xi \, \mathcal{N}(\xi) n_{\rm F}\big(\xi(2-v_{\mathrm{F}\La}/\vf)\big)\cr & \times [\delta(\xi+\La) + \delta(\xi-\La)][g^{+, -, -}_{1, 0, 0} +(g^{+, -, -}_{1, 0, 1} + g^{+, +, +}_{1, 0, 1})\xi] , 
\end{align}
where the momentum-dependent backward and forward  scattering amplitudes  have been expanded up to second order in $\xi$ following the notation introduced previously in (\ref{gxi0}-\ref{gxi20}). In the low-temperature and low-energy limits, this equation   becomes 
\begin{equation}
\partial_\ell v_{{\rm F}\ell} = \frac{1}{8}\vf\Big(\frac{V}{\pii \vf } \ee^{-2\ell} + \frac{V\mu}{\pii \vf } \ee^{-\ell}\Big).  
\end{equation}
This leads to the renormalized Fermi velocity in the low-energy limit 
\begin{equation}
   \vf^*= \vf\Big[1+ \frac{V}{8\pii \vf}\Big(\frac{1}{2}+\mu\Big)\Big].  
\end{equation} 
In weak coupling, $ \vf^* $ differs from $ \vf $ only by a few percents.

\section{Choice of the regulator}
\label{Regul}

The regulator $ r_a(x) $ is realized as a smooth step function, and depends on a rigidity parameter $a$ (numerically $ a \approx 10 $), such that $ r_{a = \infty}(x) = \Theta(x - 1) $, where $ \Theta(x) $ is the Heaviside function. Its expression is the following:
\begin{equation}
r_a(x) = g( a x - a + 1/2  ),
\end{equation}
where
\begin{equation}
    \begin{split}
g(x)  & = \dfrac{f(x)}{f(x)+f(1-x)}, \\
f(x)  & = \begin{cases}
\ee^{-1/x} \text{ if } x>0, \\
0 \text{ otherwise.}
\end{cases}
\end{split}
\end{equation}

\begin{figure}[!]
    \centering
    \includegraphics[scale=0.45]{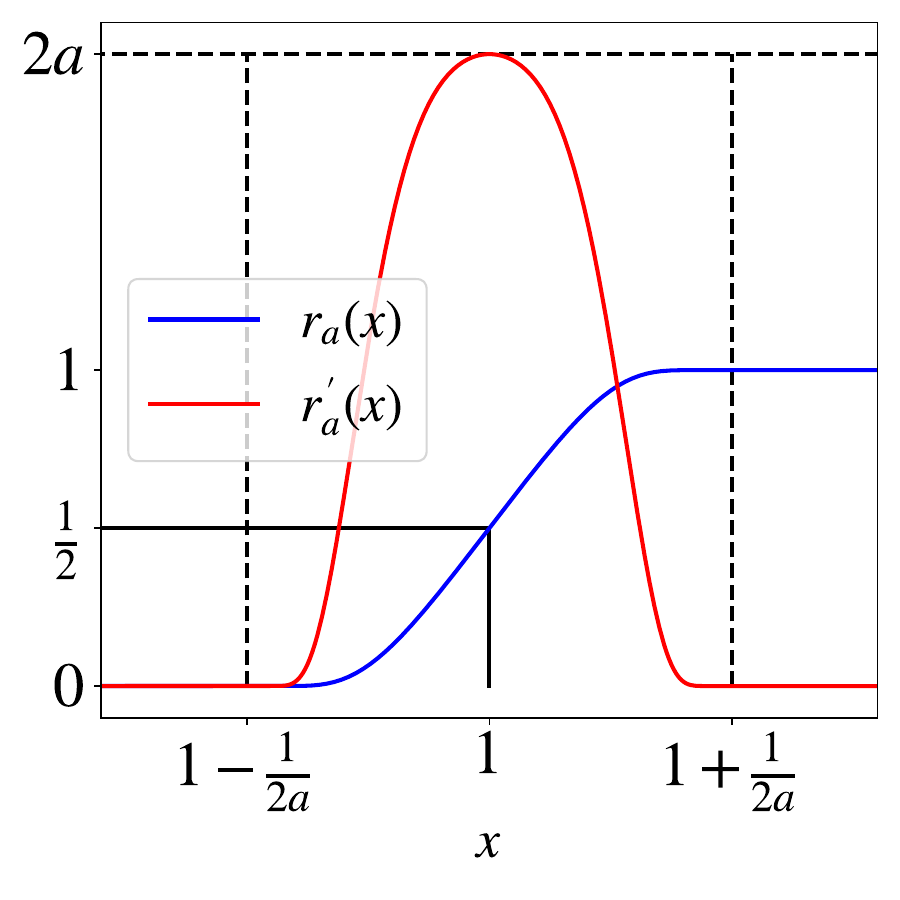}
    \caption{The regulator is such that $ r_a(0) = 0 $, $ r_a(x \gg 1) = 1 $ and $r_a(1) = 1/2$. }
    \label{Reg}
\end{figure}
The regulator is shown in \fig ~\ref{Reg}. It enters the flow equations through the function $ \theta_{\La}(k)$ in the regularized free propagator (see Eqs.~\eqref{Reg_propagator}).  This function only depends on the momentum $k$ through the variable $ \xi(k) = \xi$, and is given by
\begin{equation}
    \theta_\La(k) = r_a( | \xi | /\La).
\end{equation}
  
Such a cutoff procedure is meant to reproduce the Wilsonian RG approach\footnote{Indeed, one recovers a sharp cutoff in the limit $ a \to \infty $, that is $ r_{\infty}(|\xi|/\La) = \Theta( |\xi| - \La ) $.}, which amounts to a progressive integration of the degrees of freedom. Here, the UV degrees of freedom are integrated first, and the RG flow leads to a low-energy effective theory. In the case of one-dimensional fermions, the low-energy theory   corresponds to a model with a linear spectrum comprising two branches centered around the two Fermi points. This is of course in stark contrast to the bosonic case for which the low-energy theory is described by modes of momenta $ k\approx 0 $.

Let us now clarify the structure of the bubble $\mathcal{L}^{\mathrm{ph,pp}}_{p,q} $.  Each bubble is made of two factors: the first one is proportional to the cutoff function while the second is proportional to the derivative of this function with respect to the RG parameter $\La$ --- unslashed and slashed fermion lines respectively, in diagrams of Figs.~\ref{dgDiag}, \ref{dZDiag} and \ref{dChiDiag}.  Since the cutoff function is roughly a regularized step function, its derivative is  a regularized Dirac function, whose effect is a selection of modes of energy $ \xi \sim \La $, and hence reproduces Wilson's idea.

\begin{figure}[h!]
    \centering
    \includegraphics[scale=1.25]{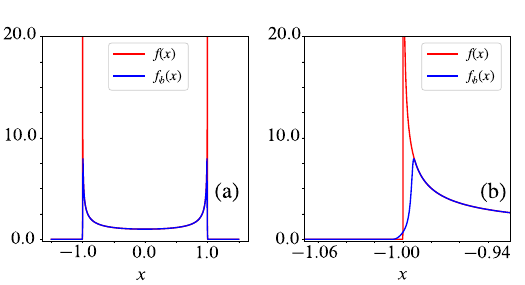}
    \caption{The van Hove singularity is regularized thanks to a smooth gate function $ G_b(x) $, and the original density of states $f(x)$ is replaced by the regularized function $ f_b(x) $ whose sharpness is controlled by the parameter $b$, with which we have $ f_{b\to \infty}(x) = f(x) $.}
    \label{RegVanHove}
\end{figure}

\paragraph*{Van Hove singularity regularization.} The regulator function $ r_b(x) $ can be used to   regularize the van Hove singularity. The density of states has the schematic form:
\begin{equation}
f(x) = \dfrac{\Theta(1 - |x|)}{\sqrt{1 - x^2}},
\end{equation}
and is singular at $ x = \pm 1 $. In order to regularize this function, we first define a regularized gate function:
\begin{equation}
G_b(x) = r_b(x+2) \big(1 - r_b(x) \big),
\end{equation}
and then make the following replacement:
\begin{equation}
f(x) \rightarrow f_b(x) = \dfrac{G_b(x)}{\sqrt{1 - x^2 G_b(x)}}.
\end{equation}
The regularized van Hove singularity is shown in Fig.~\ref{RegVanHove}. Such a regularization is advantageous, because it produces a smooth function, well suited for numerical evaluations. Furthermore, the error due to the regularization is  restricted to small segments around the singular points. This is because the regulator is built out of functions whose variation support is compact. The total number of states is recovered in the limit $b\to \infty$ (numerically $ b \approx 10^3 $).

\section{Numerical determination of phase boundaries}

Boundaries of the phase diagrams are determined using a dichotomy algorithm. The algorithm is initialized by specifying several parameters:
\begin{itemize}
    \item $ V_{-}<0 $ and $ V_{+}>0 $, with typically $ V_{+} = - V_{-} = 2 $,
    \item $ U_{-}<0 $ and $ U_{+}>0 $, with typical values $ U_{+} = - U_{-} = 2 $,
    \item a number $N$ which determines vertical lines $ U_i = U_{-} + i (U_{+} - U_{-})/N  $ for $ 0 \leqslant i \leqslant N $ ($N\approx 60$),
    \item a number $G>1$ specifying the total number of dichotomy iterations (in practice $G=10$).
\end{itemize}
In the first step, the phase is determined on each point of coordinates $ (U_i, V_{\pm}) $. Then a dichotomy is performed on each vertical line, until the final number of generations is reached. The vertical line setting is convenient because the computations done on two different lines are independent from each other, which allows the use of parallelization.
\begin{figure}[h!]
    \centering
    \includegraphics[scale=0.65]{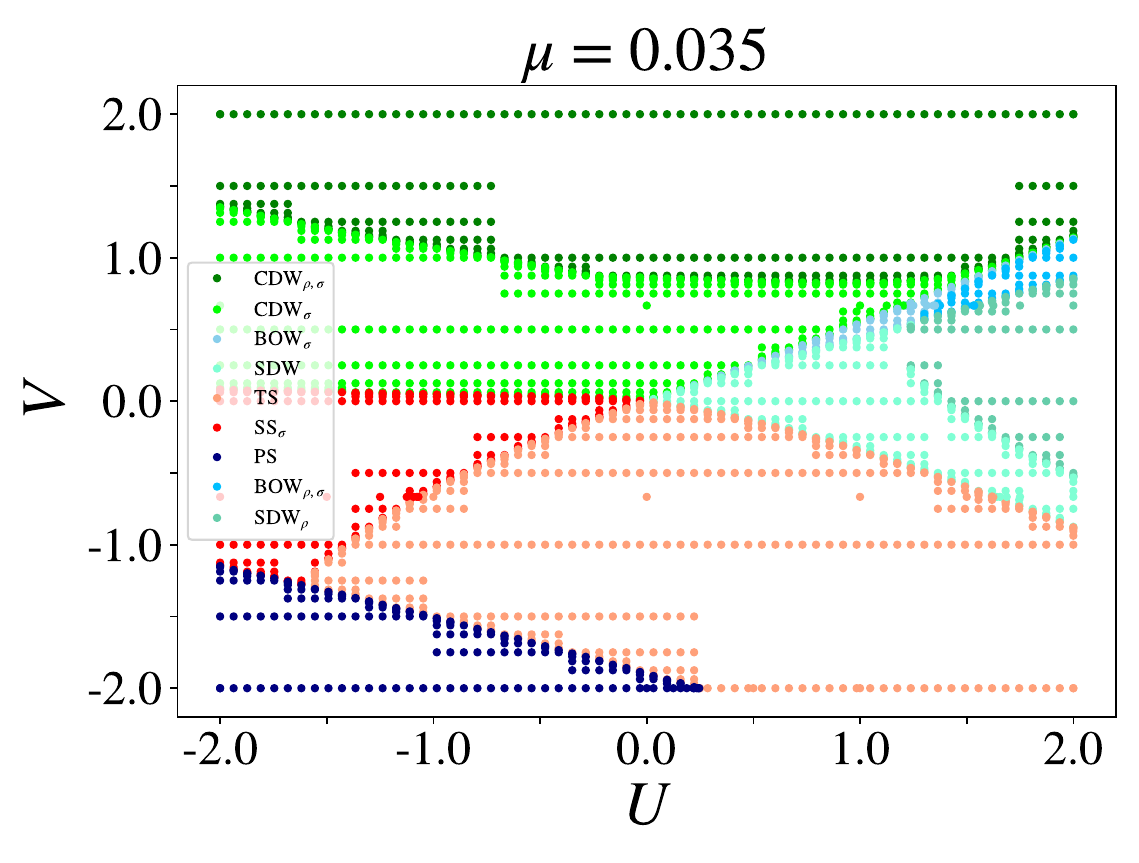}
    \caption{An example of a raw phase diagram obtained by the dichotomy algorithm explained in the text and which leads to Fig.~\ref{UVmu_0.035} with continuous boundaries.}
    \label{rawdiag}
\end{figure}
An example of a raw phase diagram is shown in Fig. \ref{rawdiag}.

\newpage

\bibliography{biblio2.bib}

\end{document}